\documentclass[fleqn,usenatbib]{mnras}

\usepackage{newtxtext,newtxmath}

\usepackage[T1]{fontenc}

\DeclareRobustCommand{\VAN}[3]{#2}
\let\VANthebibliography\thebibliography
\def\thebibliography{\DeclareRobustCommand{\VAN}[3]{##3}\VANthebibliography}

\usepackage{graphicx}	%
\usepackage{amsmath}	%
\usepackage{float}
\usepackage{lipsum}
\usepackage{xcolor}
\usepackage{natbib}
\usepackage{wasysym}
\usepackage{threeparttable}
\usepackage{tgbonum}
\usepackage{epstopdf}

\newcommand{\dd}[1]{\mathrm{d}#1}
\newcommand{\degrees}{\ensuremath{^\circ}}
\newcommand{\hours}{\ensuremath{^\mathrm{h}}}
\newcommand{\minutes}{\ensuremath{^\mathrm{m}}}

\title[Herschel-ATLAS Data Release 3]{Herschel-ATLAS Data Release III: Near-infrared counterparts in the South Galactic Pole field -- Another 100 000 submillimetre galaxies}

\author[B. A. Ward et al.]{
B. A. Ward$^{1}$\thanks{E-mail: wardb2@cardiff.ac.uk},
S. A. Eales$^{1}$,
E. Pons$^{2}$,
M. W. L. Smith$^{1}$,
R. G. McMahon$^{2}$,
L. Dunne$^{1}$,
R. J. Ivison$^{3}$,
\newauthor
S. J. Maddox$^{1}$,
M. Negrello$^{1}$
\\
$^{1}$School of Physics and Astronomy, Cardiff University, The Parade, Cardiff, CF24 3AA, UK\\
$^{2}$Institute of Astronomy, University of Cambridge, Madingley Road, Cambridge, CB3 0HA, UK\\
$^{3}$European Southern Observatory, Karl-Schwarzschild-Strasse 2, D-85748 Garching, Germany
}

\date{Accepted XXX. Received YYY; in original form ZZZ}

\pubyear{2020}

\begin{document}
\label{firstpage}
\pagerange{\pageref{firstpage}--\pageref{lastpage}}
\maketitle

\begin{abstract}
In this paper we present the third data release (DR3) of the \textit{Herschel} Astrophysical Terahertz Large Area Survey (H-ATLAS). We identify likely near-infrared counterparts to submillimetre sources in the South Galactic Pole (SGP) field using the VISTA VIKING survey. We search for the most probable counterparts within 15 arcsec of each \textit{Herschel} source using a probability measure based on the ratio between the likelihood the true counterpart is found close to the submillimetre source and the likelihood that an unrelated object is found in the same location. For 110 374 (57.0\%) sources we find galaxies on the near-infrared images where the probability that the galaxy is associated to the source is greater than 0.8. We estimate the false identification rate to be 4.8\%, with a probability that the source has an associated counterpart on the VIKING images of 0.835$\pm$0.009. We investigate the effects of gravitational lensing and present 41 (0.14\,deg$^{-2}$) candidate lensed systems with observed flux densities > 100\,mJy at 500\,$\mu$m. We include in the data release a probability that each source is gravitationally lensed and discover an additional 5 923 sources below 100\,mJy that have a probability greater than 0.94 of being gravitationally lensed. We estimate that $\sim$ 400 -- 1 000 sources have multiple true identifications in VIKING based on the similarity of redshift estimates for multiple counterparts close to a \textit{Herschel} source. The data described in this paper can be found at the H-ATLAS website.
\end{abstract}


\section{Introduction}
\label{sec:intro}

Beginning with the submillimetre Common User Bolometer Array (SCUBA; \citealt{Holland_1999}) on the James Clerk Maxwell Telescope (JCMT), a number of surveys have discovered an abundance of high-redshift, highly-luminous far-infrared (FIR) sources from their redshifted emission at submillimetre (sub-mm) wavelengths (SMGs; \citealt{Smail_1997}; \citealt{Barger_1998}; \citealt{Hughes_1998}; \citealt{Eales_1999}). Based on studies of the extragalactic background light (EBL) we now predict that there is a similar amount of light absorbed by dust and then re-emitted in the FIR and sub-mm regimes as there is seen directly in the optical and UV (\citealt{Puget_1996}; \citealt{Dwek_1998}; \citealt{Fixsen_1998}; \citealt{Dole_2006}; \citealt{Driver_2008}), suggesting that a great quantity of the star formation activity in the Universe is obscured by dust. The population of SMGs found from large blank-field surveys are found to be heavily dust-obscured and are likely to be hosts to this hidden star formation. SMGs are rare in the local Universe, but increase in number density with redshift peaking at z $\sim$ 2.5 (\citealt{Chapman_2003}; \citealt{Chapman_2005}; \citealt{Wardlow_2011}; \citealt{Casey_2014}), making them most abundant when the star formation density was at its peak. These galaxies contribute a negligible quantity to the total cosmic star formation rate density (CSFRD) in the local Universe, but form > 50\% of the CSFRD at z > 1 (\citealt{Casey_2012}). For this reason, they are often considered as the progenitors of the massive elliptical galaxies observed locally (e.g. \citealt{Toft_2014}; \citealt{Valentino_2020}) and provide insight into formation processes of galaxies at cosmic noon. Producing large samples of SMGs could be the key to understanding obscured star formation and provide the link between the early and local Universe.

The \textit{Herschel} Astrophysical Terahertz Large Area Survey (H-ATLAS; \citealt{Eales_2010}) was the largest open time submillimetre survey carried out with the \textit{Herschel} Space Observatory (\citealt{Pilbratt_2010}), covering an area of 660\,deg$^{2}$ with PACS (\citealt{Poglitsch_2010}) at 100- and 160-$\mu$m and SPIRE (\citealt{Griffin_2010}) at 250-, 350- and 500-$\mu$m. The survey area is split into three regions chosen to avoid emission from dust in the Galactic plane and to cover the best studied sky fields at high Galactic latitudes: the North Galactic Pole (NGP) region measures 180.1\,deg$^{2}$ centred at R.A. of 13\hours 18\minutes and decl. of +29\degrees 13\arcmin (J2000), which covers the Coma cluster; three fields designed to overlap with the Galaxy and Mass Assembly (GAMA) survey along the celestial equator at approximately 9\hours, 12\hours and 15\hours totalling a survey area of 161.6\,deg$^{2}$; and the South Galactic Pole (SGP) field covering 317.6\,deg$^{2}$ centred at R.A. of 0\hours 6\minutes and decl. of -32\degrees 44\arcmin (J2000) (\citealt{Smith_2017}).

The first public data release (DR1) of H-ATLAS covered the three equatorial fields. The details describing the image processing of the DR1 catalogue are found in \citealt{Valiante_2016}. The catalogue includes 120 230 sources in the five PACS and SPIRE photometric bands with 44 835 optical counterpart identifications to the \textit{Herschel} sources found in \citealt{Bourne_2016}. The second public data release (DR2) covers the NGP and SGP fields that together form $\sim$ 75\% of the entire survey. The Herschel images from this release are described in Paper I (\citealt{Smith_2017}), the sub-mm sources detected from the images are presented in Paper II (\citealt{Maddox_2018}) and the identification of sub-mm source counterparts in the NGP are described in Paper III (\citealt{Furlanetto_2018}). 

This paper describes the results of a search for the counterparts in the SGP, the largest of the H-ATLAS fields, which has coverage from both the VISTA Kilo-degree Infrared Galaxy (VIKING) survey and the Kilo Degree Survey (KiDS; \citealt{deJong_2013}) in near-infrared ($ZYJHK_s$) and optical ($ugri$) wavebands, respectively, as well as spectroscopic redshifts from the 2dF survey. A preliminary counterpart identification analysis using the Two Micron All Sky Survey (2MASS; \citealt{Skrutskie_2006}) found just 3 444 \textit{Herschel} sources in the SGP within five arcsec of a 2MASS galaxy. Here we present a comprehensive identification analysis using the likelihood ratio technique for identifying VIKING counterparts to H-ATLAS DR2 sources.

The layout of the paper is as follows. In Section \ref{sec:data} we discuss the Herschel and VIKING surveys and the method used to separate stellar contaminants from the data. Section \ref{sec:likeihood_ratio_method} describes the likelihood ratio method for identifying counterparts to sub-mm detected sources. In Section \ref{sec:results} we present the key results of the resulting crossmatched sample and describe how we obtain redshift estimates for the sub-mm sources and near-infrared counterparts. We also compare the number counts to other H-ATLAS fields. Section \ref{sec:lensed_population} outlines a method for identifying candidate lensed galaxies across the SGP at flux densities down to the 4\,$\sigma$ detection limit at 500\,$\mu$m. We show the number counts and redshift distribution of this sample and present our brightest (> 100\,mJy) gravitationally lensed candidates. Finally, we summarize our main results in section \ref{sec:conclusions}. Throughout the paper all magnitudes are given in the Vega system.

\section{Data}
\label{sec:data}
\subsection{Herschel-ATLAS Sources}
\label{sec:h-atlas_sources}

We use the H-ATLAS DR2 catalogue described in \citealt{Maddox_2018}, which contains 153 367 sources for the NGP field and 193 527 sources for the SGP detected at more than 4\,$\sigma$ significance in any of the 250-, 350- or 500-$\mu$m SPIRE bands. Sources were detected from the SPIRE maps using the MADX (\citealt{Maddox_2020}) algorithm that identifies sources from peaks in the signal to noise ratio (SNR) maps. The measurement of source fluxes is optimised using a matched filter applied to the signal and noise maps, the details of which can be found in \citealt{Valiante_2016}. 182 282 of the sources in the SGP have an SNR of greater than 4 at 250\,$\mu$m. The remaining $\sim$ 6\% of sources have a lower $\textrm{SNR}_{250}$ value, but are selected based on their 350- or 500-$\mu$m values. These objects have sub-mm colours indicating a high redshift and are the most likely candidates for having erroneously matched counterparts resulting from chance alignments or lensing (\citealt{Negrello_2010}; \citealt{Bourne_2014}). These objects could therefore be removed prior to our counterpart analysis to improve the false identification rate of the sample, but would impact on our search for gravitationally lensed sources and so are kept in our catalogue. Note that the sources extracted from the maps only come from images that are made from more than one individual \textit{Herschel} observation, which means that the area covered by the catalogues is slightly smaller than the size of the maps. The DR2 catalogue covers an area of 177\,deg$^2$ and 303\,deg$^2$ in the NGP and SGP fields, respectively. Given that the catalogue described in \citealt{Maddox_2018} came from maps that have at least two observations, the fields were searched for moving objects. In the first data release of the equatorial GAMA fields nine asteroids were detected, but no moving sources were found in either the NGP or SGP.

The 4\,$\sigma$ flux density limits for the three SPIRE bands are 29.4\,mJy, 37.4\,mJy and 40.6\,mJy for 250, 350 and 500 microns, respectively (\citealt{Valiante_2016}). Flux bias, the phenomenon where measured flux densities are found to be systematically higher than the true flux densities, is a common problem in submillimetre surveys. At the 4\,$\sigma$ detection level, the measured flux densities are expected to be approximately 20\%, 5\% and 4\% higher than the true values, respectively (\citealt{Maddox_2018}).

\subsection{VISTA VIKING Counterparts}
\label{sec:vista_viking_counterparts}

The Visible and Infrared Survey Telescope for Astronomy (VISTA) is a 4-m wide field survey telescope located at the European Southern Observatory's (ESO) Paranal observatory (\citealt{Emerson_2010}). One of the public surveys of VISTA is VIKING, imaging a total of 1 500\,deg$^2$ across the celestial equator and southern Galactic cap in five filters: $Z$, $Y$, $J$, $H$ and $K_s$, covering this area to a 5\,$\sigma$ depth of 23.1, 22.3, 22.1, 21.5 and 21.2 at these bands, respectively (\citealt{Edge_2013}). The area of the VIKING survey covering the southern Galactic cap spans from RA 22\hours to 3.5\hours with a width of 10 degrees. The H-ATLAS fields were chosen to maximize the coverage with complementary multi-wavelength data, including an overlap with the VIKING survey of more than 360 square degrees. 

For our object catalogue we use the fourth data release of VIKING to extract all objects to within 15 arcsec of the 250-$\mu$m position of each \textit{Herschel} source. The counterpart analysis of the 9-h GAMA field by \citealt{Fleuren_2012}, covering approximately 54 square degrees, also used VIKING near-infrared data and found that they were able to reliably crossmatch 51\% of all 250\,$\mu$m \textit{Herschel} sources. This is a substantial increase from the percentage found during analyses in the optical $r$-band that typically reach 35 -- 40\%. We expect a similar percentage of reliably matched sources. Given the large survey area of the SGP and the higher percentage of reliable matches found during near-infrared counterpart analyses, we present here the largest sample of reliably matched \textit{Herschel} sources, exceeding the combined number found to date.

\subsection{Star-Galaxy Classifier}
\label{sec:star_galaxy_classifier}

\begin{figure}
	\includegraphics[width=\columnwidth]{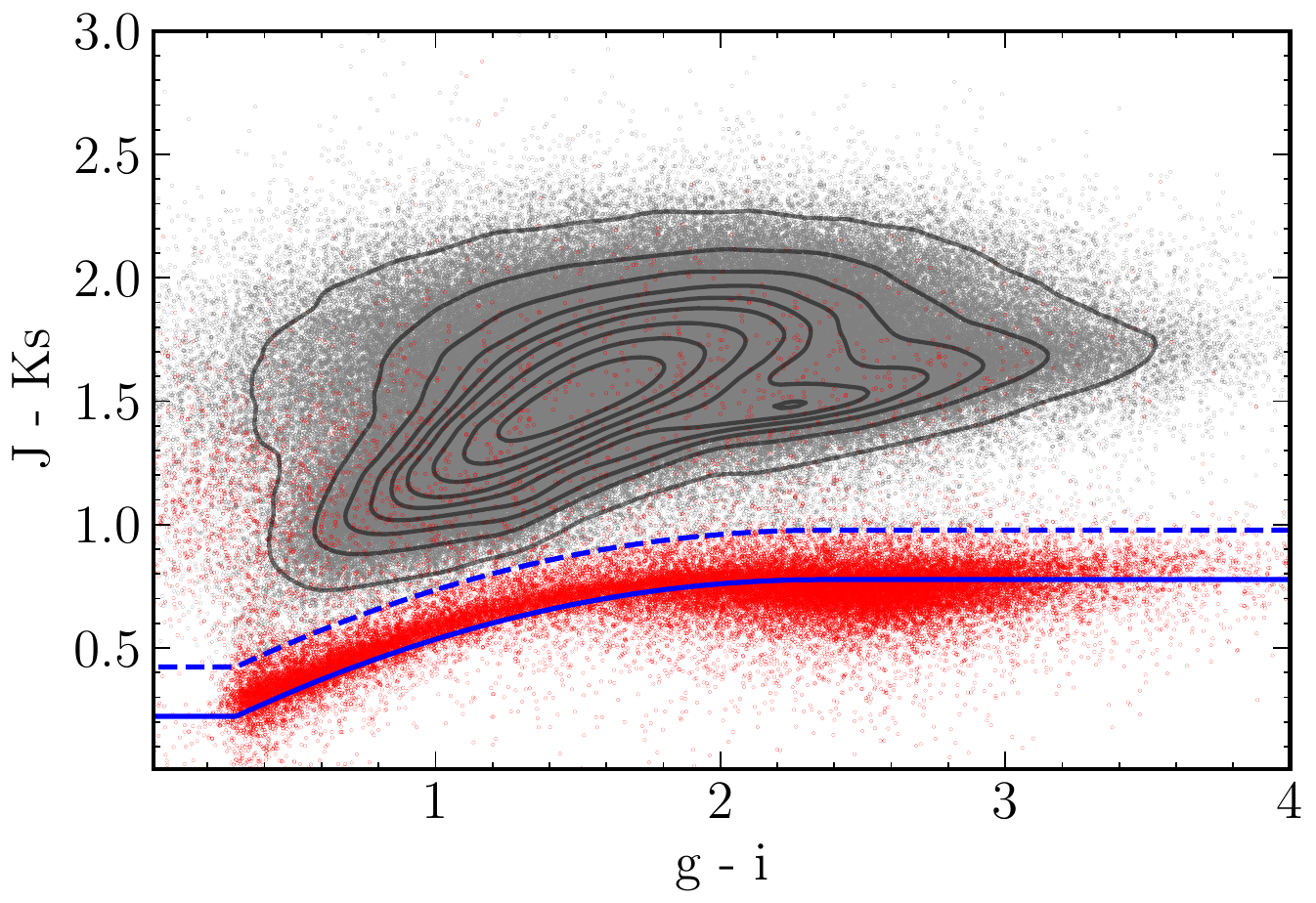}
	\caption{The colour-colour diagram of VIKING objects with KiDS identifications. The solid blue line represents the stellar locus given by equation (\ref{eq:stellar_locus}). The dashed blue line is +0.2 offset in $J - K_s$ from the stellar locus and represents the separation line between galaxies and stars. Galaxies identified using our classification process are illustrated as grey points while stars are shown as red points.}
	\label{fig:star_galaxy_classifer}
\end{figure}

A significant number of objects in the VIKING survey are stars that contaminate our counterpart analysis. In \textit{Herschel}'s Science Demonstration Phase (SDP) (\citealt{Fleuren_2012}), DR1 (\citealt{Bourne_2016}) and DR2 (\citealt{Furlanetto_2018}) counterpart analyses, a star-galaxy classifier is defined based on a combination of shape and colour parameters. These classifiers depend on the $J - K_s$ versus $g - i$ colour space by virtue of the optical bands of the Sloan Digital Sky Survey (SDSS; \citealt{Abazajian_2009}). Without SDSS coverage in the SGP field, we instead match the objects in our VIKING catalogue to the fourth data release of KiDS using the nearest neighbour within 0.5 arcsec to obtain $g - i$ colours. We follow a similar classification method to the above papers, which in turn are slightly modified versions of the method described in \citealt{Baldry_2010}.

Firstly, we classify as stellar any object with \textit{pStar}, a probability that the object is a star based on a VIKING shape parameter, greater than 0.95. This immediately classifies 51 508 objects as stellar. Failing this, we consider the $J - K_s$ versus $g - i$ colour space as shown in Fig \ref{fig:star_galaxy_classifer}. The stellar locus as decsribed in \citealt{Baldry_2010} is converted to the Vega system assuming $J_{\textrm{Vega}} = J_{\textrm{AB}} - 0.91$ and $K_{s \textrm{Vega}} = K_{s \textrm{AB}} - 1.85$:

\begin{equation}
\label{eq:stellar_locus}
f_{\textrm{locus}} = 
\begin{cases} 
	0.2228 & x < 0.3 \\
	0.05 + 0.615x - 0.13x^2 & 0.3 \leq x < 2.3 \\
	0.7768 & x \geq 2.3
\end{cases}
\end{equation}

\noindent where $x = g-i$ and is shown as the blue solid line in Fig \ref{fig:star_galaxy_classifer}. The location of the separation line beween stars and galaxies is then chosen to be $+0.2$ offset in $J - K_s$ from the stellar locus. A further 411 463 objects are classified in this way, 299 525 being extragalactic and 111 938 stellar. The remaining objects without a match in the KiDS survey are next classified based on their $J - K_s$ values. As explained in \citealt{Fleuren_2012}, objects without $g - i$ colour must lie above or below the separation line if their $J - K_s$ value is greater or less than 0.98 and 0.42, respectively. The stellar contamination for objects with $J - K_s$ above 0.98 or galaxy contamination for objects with $J - K_s$ less than 0.42 is negligible. We therefore classify an additional 102 265 objects as galaxies if they have $J - K_s$ > 0.98 and classify an additional 275 objects as stars if their value of $J - K_s$ is less than 0.42. Finally, we return to the \textit{pStar} stellar probability value and classify as stars any remaining objects that have a value of \textit{pStar} greater than 0.7. All remaining objects are classified as galaxies.

We include in the data release a flag that indicates how each object has been classified. In descending order of choice:

\begin{itemize}
	\item 0 - Stellar : \textit{pStar} > 0.95.
	\item 1 - Galaxy: $J - K_s$/$g - i$ colour plane.
	\item 2 - Stellar: $J - K_s$/$g - i$ colour plane.
	\item 3 - Galaxy: $J - K_s$ > 0.98.
	\item 4 - Stellar: $J - K_s$ < 0.42.
	\item 5 - Stellar: \textit{pStar} > 0.70.
	\item 6 - Galaxy: All remaining objects.
\end{itemize}

This method leads to the identification of 793 331 (78.9\%) extragalactic objects and 212 028 (21.1\%) stars.

\section{Likelihood Ratio Method}
\label{sec:likeihood_ratio_method}

The likelihood ratio (LR) method is used to identify the most reliable near-infrared counterparts to the 250\,$\mu$m sources from the SGP field. This method, developed by \citealt{Sutherland_1992}, assigns to all counterparts a probability of being the true identification defined by the ratio of two likelihoods: the likelihood of the true counterpart being observed at a magnitude $m$ and separation $r$ from the source, and the likelihood of a background object being found with the same properties. The LR for each possible counterpart is defined as

\begin{equation}
\label{eq:likelihood}
L = \frac{q(m, c) f(r)}{n(m, c)},
\end{equation}

\noindent where $q(m, c)$ represents the probability distribution of true counterparts with magnitude $m$ and class $c$ (e.g. any additional properties allowing us to distinguish between counterparts, such as stellar or extragalactic classes), $n(m, c)$ is the surface density distribution of background objects and $f(r)$ represents the probability distribution of the source positional errors.

The probability that an object at a distance $r$ from the source with magnitude $m$ is the true ID is then given by the reliability:

\begin{equation}
\label{eq:reliability_single}
R = \frac{L}{L + 1}.
\end{equation}

As shown in \citealt{Sutherland_1992}, this reliability assumes a single candidate with likelihood, L, and assumes no prior knowledge about the presence of other possible candidates to the same source. We find that 179 096 (92.5\%) of the 250\,$\mu$m sources in the SGP have multiple VIKING objects within 15 arcsec. For a source with multiple possible counterparts, the reliability $R_j$ that the $j$th candidate within the search radius is the true ID is then

\begin{equation}
\label{eq:reliability_multi}
R_j = \frac{L_j}{\sum_i L_i + (1 - Q)},
\end{equation}

\noindent where $i$ extends to the number of candidates found within the search radius. The value $Q$ represents the fraction of all true counterparts that are brighter than the limiting magnitude of the matching survey, thus the $(1 - Q)$ term accounts for the probability that the counterpart is not detected in VIKING. Previous studies of other H-ATLAS fields have estimated the value of $Q$ (hereafter an estimate of $Q$ shall be denoted $Q_0$), finding a wide range of values as a result of the different depths and passbands of the matching catalogues. \citealt{Furlanetto_2018} quote a value of $0.836 \pm 0.001$ for their near-infrared counterpart analysis of $\sim$ 25\,deg$^{2}$ of the NGP using K-band data obtained from the United Kingdom InfraRed Telescope (UKIRT; \citealt{Casali_2007}). A value of $Q_0 = 0.538 \pm 0.001$ is also given by \citealt{Furlanetto_2018} for an optical counterpart analysis on the whole NGP area using SDSS. Other values include $0.539 \pm 0.001$ by \citealt{Bourne_2016} for the analysis of the three equatorial GAMA fields using SDSS and $0.7342 \pm 0.0257$ by \citealt{Fleuren_2012} in a pilot analysis of the 9-h GAMA field using VIKING. It is worth noting that the value of $Q_0$ is higher in the near-infrared surveys, highlighting their greater depth.

\subsection{True Counterparts Distribution}
\label{sec:true_counterparts_distribution}

\begin{figure}
	\includegraphics[width=\columnwidth]{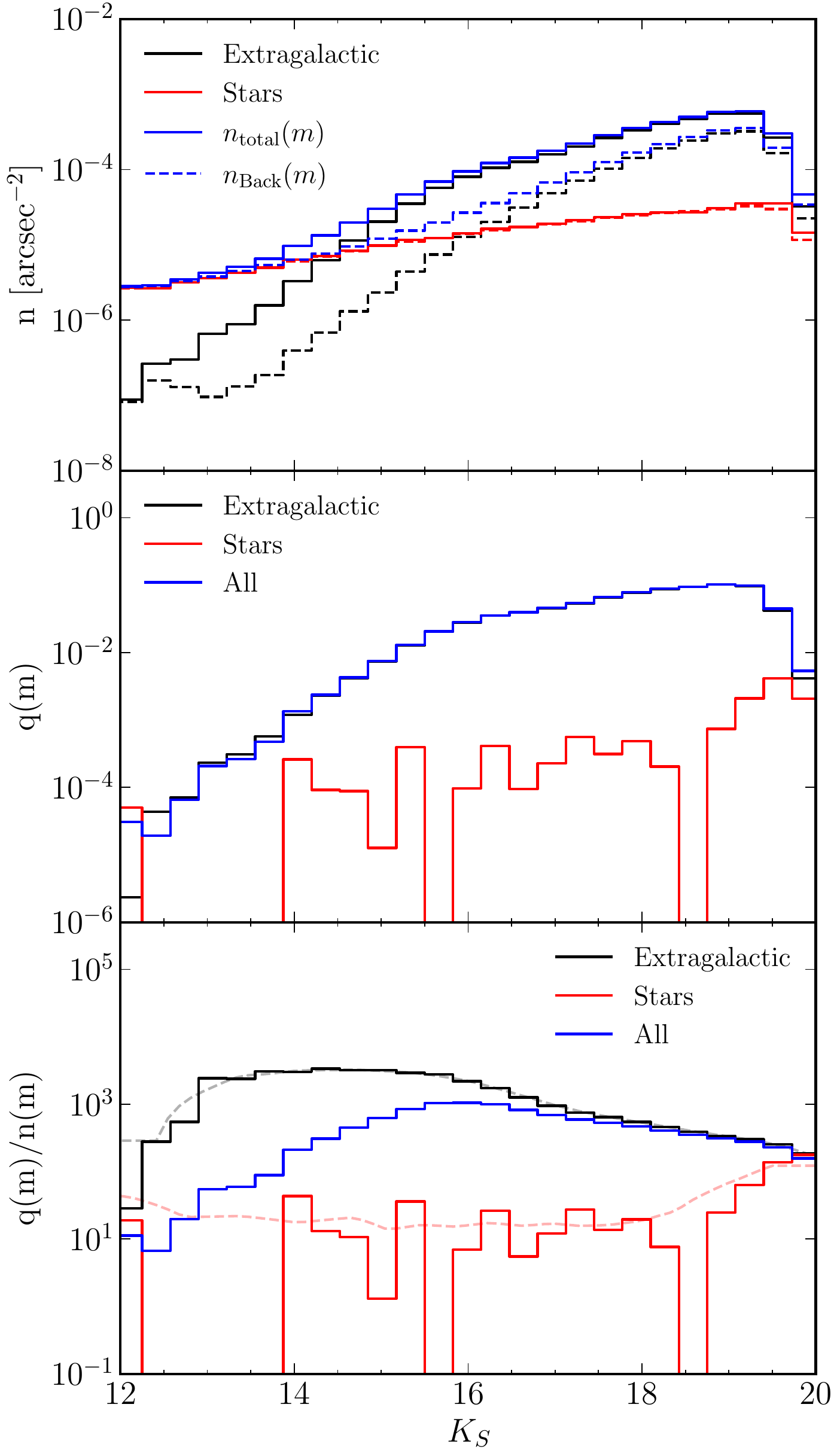}
    \caption{The $K_\textrm{s}$ magnitude distributions used to identify true counterparts. Top panel: The magnitude distribution of objects within 15 arcsec of a known position (blue line). This is divided into two classes; extragalactic (black line) and stars (red line). The background magnitude distribution is represented using dashed lines following the same colour convention. The distributions are normalized to their respective search areas, given in units of [arcsec$^{-2}$]. Middle panel: The true counterparts distribution $q(m)$ found by the subtraction of the background magnitude distribution from the total distribution. Bottom panel: The ratio of the true counterparts distribution to that of the background objects, as in equation (\ref{eq:likelihood}). The dashed lines represent the smoothed fits to $q(m)/n(m)$ for extragalactic objects and stars, used when the true value of $q(m)/n(m)$ becomes negative.}
    \label{fig:q_n_distributions}
\end{figure}

We determine the probability distribution of true counterparts, $q(m)$, using the method described in \citealt{Ciliegi_2003}. For both galaxies and stars we generate a magnitude distribution of objects found in the near-infrared images within a search radius of 15 arcsec around each \textit{Herschel} source. This forms the distribution $n^{\prime}_{\textrm{total}}$ (we denote counts that are not normalized to their search area with prime notation), from which we background subtract the counts that are found around random positions, $n^{\prime}_{\textrm{back}}$. A set of 844 715 random positions are used to sample the background count level, from which 2 917 214 objects are found within 15 arcsec. The magnitude distribution of all true counterparts, $n_{\textrm{real}}(m)$, is then given by

\begin{equation}
\label{eq:real_distribution}
n_{\textrm{real}}(m) = n^{\prime}_{\textrm{total}} - n^{\prime}_{\textrm{back}} \frac{N_{\textrm{total}}}{N_{\textrm{back}}}
\end{equation}

\noindent where the counts have been normalized to the same search area. $N_{\textrm{total}}$ and $N_{\textrm{back}}$ represent the number of \textit{Herschel} 250\,$\mu$m and background positions, respectively. The $q(m)$ distribution is then derived by normalizing $n_{\textrm{real}}(m)$ and scaling by the value $Q_0$. This normalization ensures that the integral of $q(m)$ over all magnitudes up to the limiting magnitude of VIKING is equal to the probability that a source is detected within the survey (i.e. $\int^{m_{\textrm{lim}}} q(m) \dd{m} = Q$), thus

\begin{equation}
\label{eq:q_distribution}
q(m) = \frac{n_\textrm{real}(m)}{\sum_{m_i}n_\textrm{real}(m_i)}\times Q_0,
\end{equation}

\noindent where we sum over bins of magnitude. The true counterparts distribution can be seen in the middle panel of Fig \ref{fig:q_n_distributions}. The figure shows that the $q(m)$ and $n(m)$ distributions are vastly different for galaxies and stars, highlighting the need to implement the LR analysis separately for the two classes to avoid a contamination of stars. The small percentage of stars and the poor sampling at particular magnitudes means that not all bins are well populated. In the cases where there is a small number of sources and the Poisson noise is large for a given magnitude bin, the value of $q(m)/n(m)$ becomes negative. This would not be consistent with our definitions of the likelihood ratio and reliability which requires that $L \in [0, \infty]$ such that $R \in [0, 1]$. To avoid this, we generate a smooth fit of $q(m)/n(m)$ for galaxies and stars and interpolate the missing values. We create the smooth fits using a moving average of the positive $q(m)/n(m)$ values. The window size used for the stellar distribution was made to be larger than that of the galaxies as the stellar $q(m)/n(m)$ does not change considerably  and should not be weighted highly towards over or under populated bins. The fitted functions used for the LR method are shown as dashed lines in the bottom panel of Fig \ref{fig:q_n_distributions}.

\subsection{Estimation of Q}
\label{sec:estimation_of_Q}

We generate an estimate for the probability of finding a genuine counterpart above the limiting magnitude of VIKING using the method described in \citealt{Fleuren_2012}. This method measures the value $1 - Q_0$ by counting the number of blanks (sources that have no counterpart) as a function of the search radius from the source. This is an improvement on methods determining $Q_0$ directly as it avoids the possibility of multicounting due to source clustering and/or sources with multiple genuine counterparts that would otherwise overestimate the value. 

A blank is most likely to be found in cases where the genuine counterpart is fainter than the limiting magnitude, but may also be counted as such if the counterpart lies outside the search radius or the blank is the location of a spurious SPIRE detection. To determine the number of true blanks, $B_\textrm{t}$, we consider that the number of observed blanks, $B_\textrm{obs}$, is simply the true number reduced by the number of positions that contain random interlopers, $N_\textrm{rand}$, such that

\begin{equation}
B_\textrm{obs} = B_\textrm{t} - N_\textrm{rand}.
\end{equation}

The number of interlopers can be determined from reusing the catalogue of background positions. We can rewrite $N_\textrm{rand}$ as $B_\textrm{t} \times f_\textrm{rand}$, where $f_\textrm{rand}$ represents the fraction of positions that have a random interloper. This fraction can be estimated directly from the fraction of non-blanks in the background positions:

\begin{equation}
f_\textrm{rand} = \frac{N_\textrm{back} - B_\textrm{back}}{N_\textrm{back}} = 1 - \frac{B_\textrm{back}}{N_\textrm{back}}
\end{equation}

\noindent where $B_\textrm{back}$ and $N_\textrm{back}$ are the number of blanks and total positions in the background catalogue, respectively. These values are then scaled such that $N_\textrm{back}$ is equal to the number of SPIRE sources. The observed number of blanks surrounding \textit{Herschel} positions can then be written as 

\begin{equation}
B_\textrm{obs} = B_\textrm{t} - B_\textrm{t}\times f_\textrm{rand} = B_\textrm{t} - B_\textrm{t}\times\Bigg(1 - \frac{B_\textrm{back}}{N_\textrm{back}}\Bigg) = B_\textrm{t}\times\frac{B_\textrm{back}}{N_\textrm{back}},
\end{equation}

\noindent which rearranged gives

\begin{equation}
B_\textrm{t} = \frac{N_\textrm{back}\times B_\textrm{obs}}{B_\textrm{back}}.
\end{equation}

\begin{figure}
	\includegraphics[width=\columnwidth]{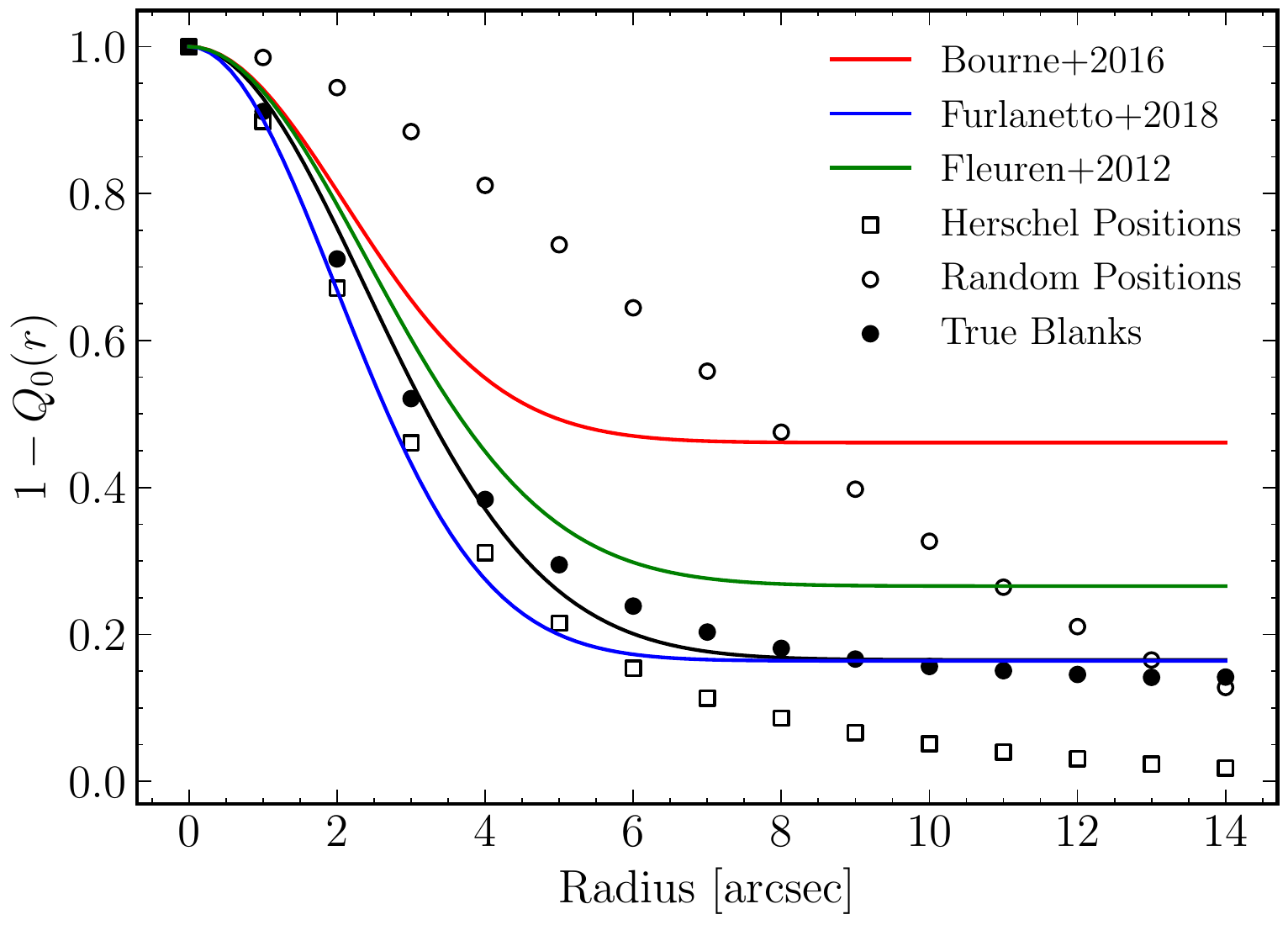}
	\caption{The method used to estimate $1 - Q_0$ by counting blanks (sources without VIKING candidates within a search radius $r$) as a function of the search radius. Open circles represent the number of blank background positions, open squares represent the number of blank \textit{Herschel} positions and the filled circles are the result of dividing the number of blank \textit{Herschel} positions by the number of blank background positions. The black line is the best fit to the points using the model described by equation (\ref{eq:blanks}). This fit yields a value of $Q_0 = 0.835 \pm 0.009$ for all candidates. The same method considering just extragalactic objects gives a value of $Q_0 = 0.823 \pm 0.009$. The values estimated by \protect\citealt{Fleuren_2012}, \protect\citealt{Bourne_2016}  and \protect\citealt{Furlanetto_2018} (near-infrared counterpart analysis) are plotted as green, red and blue lines, respectively.}
	\label{fig:Q0}
\end{figure}

By dividing by $N_\textrm{back}$ we find the fraction of \textit{Herschel} sources that are true blanks,

\begin{equation}
\label{eq:true_blanks_N}
\frac{B_\textrm{t}}{N_\textrm{back}} = \frac{B_\textrm{obs}}{B_\textrm{back}},
\end{equation}

\noindent which becomes our estimate of $1 - Q_0$. Thus we need only divide the number of observed blanks surrounding \textit{Herschel} sources by the number of blanks found at random positions. Our estimate of $1 - Q_0$ is plotted as a function of search radius in Fig \ref{fig:Q0}. We can model the dependence of the true blanks on the search radius as 

\begin{equation}
\label{eq:blanks}
B(r) = 1 - Q_0F(r),
\end{equation}

\noindent where

\begin{equation}
F(r) = \int_0^r 2\pi r^{\prime} f(r^{\prime}) \dd{r^{\prime}}.
\end{equation}

For the positional offset distribution, $f(r)$, we choose a Gaussian profile based on the assumption that H-ATLAS sources are point-like and that the positional errors are symmetric in both RA and Dec. Assuming a Gaussian distribution with width $\sigma_\textrm{pos}$, the function $F(r)$ becomes

\begin{equation}
\label{eq:F(r)}
F(r) = 1 - e^{-\frac{r^2}{2\sigma_{\textrm{pos}}^2}}.
\end{equation}

In fitting the function given in equation (\ref{eq:blanks}) to our estimate for the fraction of true blanks, we find a value for $Q_0$ of $0.835 \pm 0.009$. When considering only galaxies we find $Q_0 = 0.823 \pm 0.009$. The small difference between the two values comes from the minimal number of stars that could be identified and removed based upon the method in Section \ref{sec:star_galaxy_classifier}. We note that our value of $Q_0$ closely matches the value obtained by \citealt{Furlanetto_2018} and their near-infrared analysis of the NGP. 

In Fig \ref{fig:Q0} we see the similarity between our derived $B(r)$ function and that of \citealt{Furlanetto_2018}. The large separation between our $B(r)$ function (as well as those of \citealt{Furlanetto_2018} and \citealt{Fleuren_2012}) and the function derived by \citealt{Bourne_2016} highlights the difference in using near-infrared surveys rather than optical for the matching survey.

The fitting function used to determine $Q_0$ also provides an estimate for the value of the positional error which we measure to be $\sigma_{\textrm{pos}} = 2.388 \pm 0.065$. The fact that our Gaussian profile becomes flat and near identical to \citealt{Furlanetto_2018} above eight arcsec suggests that there is little benefit to increasing the search radius beyond $\sim$ 3\,$\sigma_{\textrm{pos}}$ to increase the number of counterparts found. An accurate value for the positional uncertainty is important since this enters the likelihood ratio formula above in the form of the positional offset distribution $f(r)$. Given that the likelihood ratio is calculated for all candidates found in the 250\,$\mu$m maps, we now describe the method we use to estimate $\sigma_{\textrm{pos}}$ for each counterpart.

\subsection{Positional Uncertainty}
\label{sec:positional_uncertainty}

To determine the 1\,$\sigma$ positional uncertainty of the 250\,$\mu$m \textit{Herschel}-ATLAS detections, we implement an empirical alteration to the theoretical form derived by \citealt{Ivison_2007}. If the width $\sigma_{\textrm{pos}}$ of $f(r)$ is simply the SPIRE positional error (neglecting the smaller positional error from the matching catalogue) then it is shown that the positional uncertainty is dependent on the signal-to-noise ratio (SNR) and the full-width at half-maximum (FWHM) of detections according to

\begin{equation}
\label{eq:pos_uncertainty_theor}
\sigma_{\textrm{pos}} = 0.6 \times \frac{\textrm{FWHM}}{\textrm{SNR}}.
\end{equation}

In previous H-ATLAS papers (\citealt{Smith_2011}; \citealt{Bourne_2014}; \citealt{Bourne_2016}; \citealt{Furlanetto_2018}) the value of $\sigma_{\textrm{pos}}$ is estimated by deriving histograms of the separations between the positions in the SPIRE catalogue and all objects in the matching catalogue within a large radius (typically around 50 arcsec). The two dimensional histogram of the separation in RA and Dec is then modelled using a three-part distribution (see \citealt{Smith_2011} for further details) to obtain a value for $\sigma_{\textrm{pos}}$. Here, an estimate for the positional uncertainty is made for each source individually using an empirical form of equation (\ref{eq:pos_uncertainty_theor}) given by

\begin{equation}
\label{eq:pos_uncertainty_k}
\sigma_{\textrm{pos}} = k \times \frac{\textrm{FWHM}}{\textrm{SNR}},
\end{equation}

\noindent where $k$ is a constant to be found. The value of $k$ is calculated by taking the 250\,$\mu$m SNR and FWHM ($\sim$ 18 arcsec) for each detection and the value of $\sigma_{\textrm{pos}}$ obtained from the fitting of blank sources in Section \ref{sec:estimation_of_Q}. From this set of $k$ values we take the median value to find an estimate of k to be 0.66. This value is used to determine an individual positional uncertainty and thus $f(r)$ for each object. Our simplified approach is a good approximation to the histogram method. \citealt{Furlanetto_2018} find that the empirical dependence of $\sigma_{\textrm{pos}}$ on SNR closely matches the theoretical form of \citealt{Ivison_2007}. Given that our method provides a similar value of $k$ we would not expect there to be any significant differences in the LR calculation between this analysis and the aforementioned papers.

\section{Results}
\label{sec:results}

\begin{figure}
	\includegraphics[width=\columnwidth]{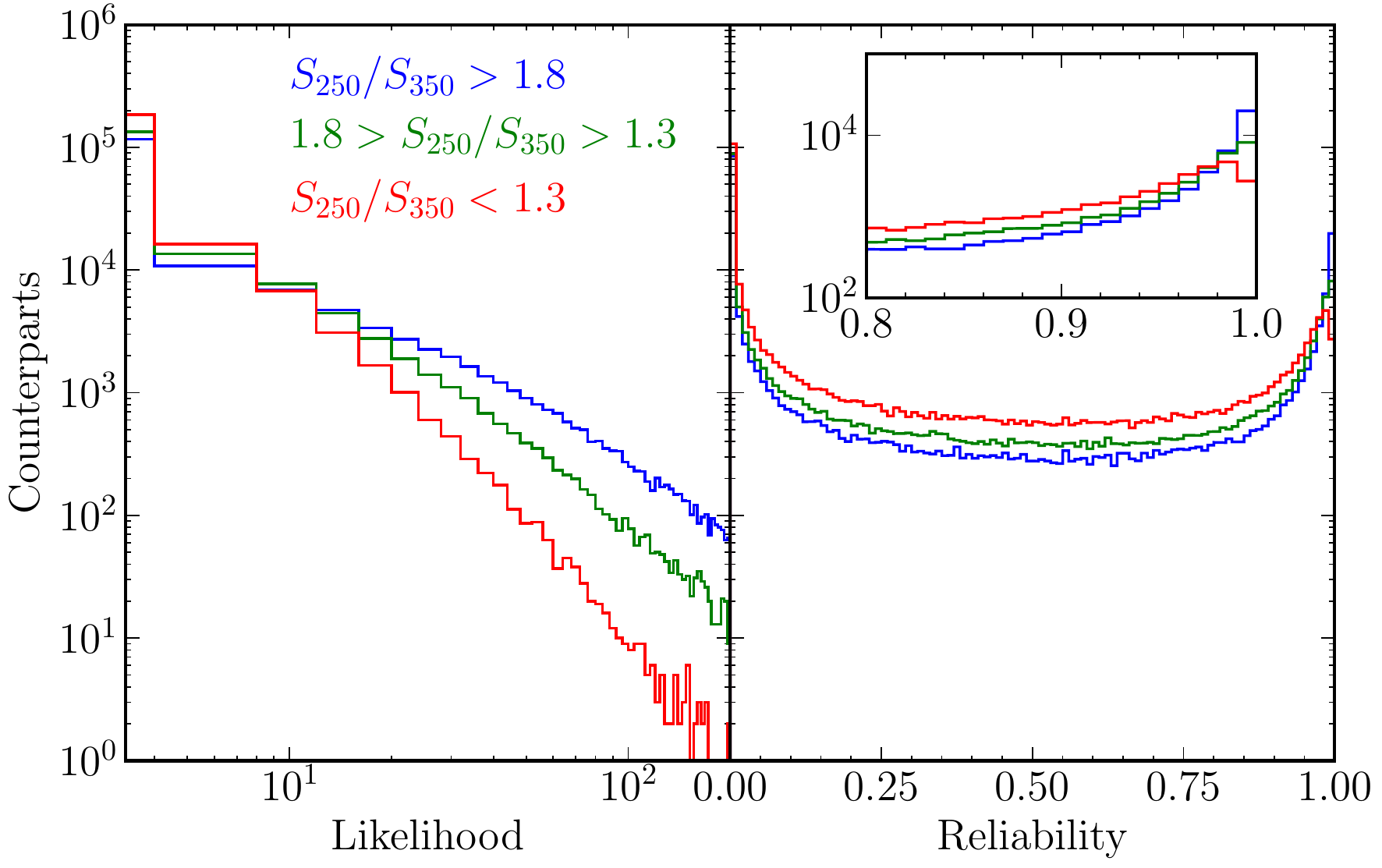}
	\caption{Likelihood ratios and reliabilities of the near-infrared counterparts to $\textrm{SNR}_{250}$ > 4 \textit{Herschel} sources separated into three bins of SPIRE colour. The inset figure shows the number of counterparts per SPIRE source as a function of reliability for the most reliable (R $\geq$ 0.8) counterparts.}
	\label{fig:lr_reliability}
\end{figure}

The LR and reliability of all possible matches within 15 arcsec of each SPIRE source was calculated using equations (\ref{eq:likelihood}) and (\ref{eq:reliability_multi}). The SPIRE catalogue contains 193 527 sources of which 190 788 have at least one possible counterpart within the search radius. Within this sample, 180 030 are sources with $\textrm{SNR}_{250} \geq 4$, 181 373 (95.1\%) are classified as extragalactic objects and 9 415 (4.9\%) as stars. In Fig \ref{fig:lr_reliability} we show the LRs and reliabilities of all potential counterparts in three bins of $S_{250}/S_{350}$ colour. It was shown by \citealt{Bourne_2016} and \citealt{Furlanetto_2018} that the fraction of sources with high LR and reliability counterparts is higher for blue \textit{Herschel} sources ($S_{250}/S_{350} > 1.8$) than it is for red sources ($S_{250}/S_{350} < 1.3$), likely due to red sources having a higher probability of lying at high redshifts and therefore have a lower probability of being detected (see their Figures 8 and 5, respectively). In comparison, our red sources constitute a slightly greater fraction of the sources with the highest reliabilities (see inset). This shows that we have less of a dependence on blue sources for reliable matches, a likely result of the deeper VIKING images. Fig \ref{fig:lr_reliability} also shows that SPIRE sources of all sub-mm colours have a large number of high and low reliability matches. A substantial number of the low reliability counterparts are due to chance alignments. The results of the LR method are summarized in Table \ref{tab:lr_results}. 

\begin{table}
	\centering
	\caption{Results of the likelihood ratio method.}
	\label{tab:lr_results}
	\begin{tabular}{lcr}
		\hline
		\hline
		H-ATLAS sample &  & \\
		\hline
		SPIRE & 193 527 & \\
		$N_{\textrm{match}} \geq 1$ & 190 788 & \\
		($N_{\textrm{match}} \geq 1$) \& ($\textrm{SNR}_{250} > 4$) & 180 030 & (94.4\% of $N_{\textrm{match}} \geq 1$)\\
		Extragalactic & 181 373 & (95.1\% of $N_{\textrm{match}} \geq 1$)\\
		Stars & 9 415 & (4.9\% of $N_{\textrm{match}} \geq 1$) \\
		Reliable IDs (R $\geq$ 0.8) & 111 065 & \\
		Reliable Extragalactic IDs & 110 374 & (99.4\% of reliable sample) \\
		Reliable Star IDs & 691 & (0.6\% of reliable sample) \\
		\hline
		\hline
	\end{tabular}
\end{table}

\subsection{Reliable ID Sample}
\label{sec:reliable_id_sample}

We choose to define reliable counterparts as those matches with R $\geq$ 0.8 in order to minimize the number of false counterparts and ensure a sample that is dominated by sources with low blending. We identified 111 065 reliable counterparts, which means that 58.2 per cent of the SPIRE sources that have at least one possible counterpart were matched with a high reliability. Of the reliable IDs, 110 374 (99.4\%) are classified as galaxies and just 691 (0.6\%) as stars. The increase in the fraction of galaxies from the total sample suggests that the LR method is biased against stars as intended.

With the LR method we are still expected to erroneously match certain objects. Assuming that the probability of a match being falsely made is given by $1 - R$ of each counterpart, then we can estimate the number of false IDs with $R \geq 0.8$ using

\begin{equation}
N_{\textrm{False}} = \sum_{R \geq 0.8} (1 - R).
\end{equation}

From this we predict 5 343 sources to be falsely labelled as reliable. This represents 4.8 per cent of all reliable matches and is compared to 4.7 per cent in the GAMA fields, 4.5 per cent for the NGP (near-infrared analysis) and 4.2 per cent for the SDP.

The completeness of the reliable sample is defined as the fraction of 250\,$\mu$m sources with a counterpart that are recovered with $R \geq 0.8$ (\citealt{Smith_2011}) and is calculated as

\begin{equation}
\label{eq:completeness}
\eta = \frac{n(R \geq 0.8)}{n(\textrm{SNR}_{250} \geq 4)}\frac{1}{Q_0},
\end{equation}

\noindent where $n(\textrm{SNR}_{250} \geq 4)$ estimates the true number of sources. Note that a value of $\eta = 1$ means that the fraction of reliable counterparts reaches a maximum value at $Q_0$ as expected. For the reliable extragalactic counterparts we obtain $\eta = 78.0$ per cent which is close to the completeness of 74 per cent found by \citealt{Furlanetto_2018} and similar to the completeness in the GAMA fields (73.0 per cent) obtained by \citealt{Bourne_2016}.

\begin{figure}
	\includegraphics[width=\columnwidth]{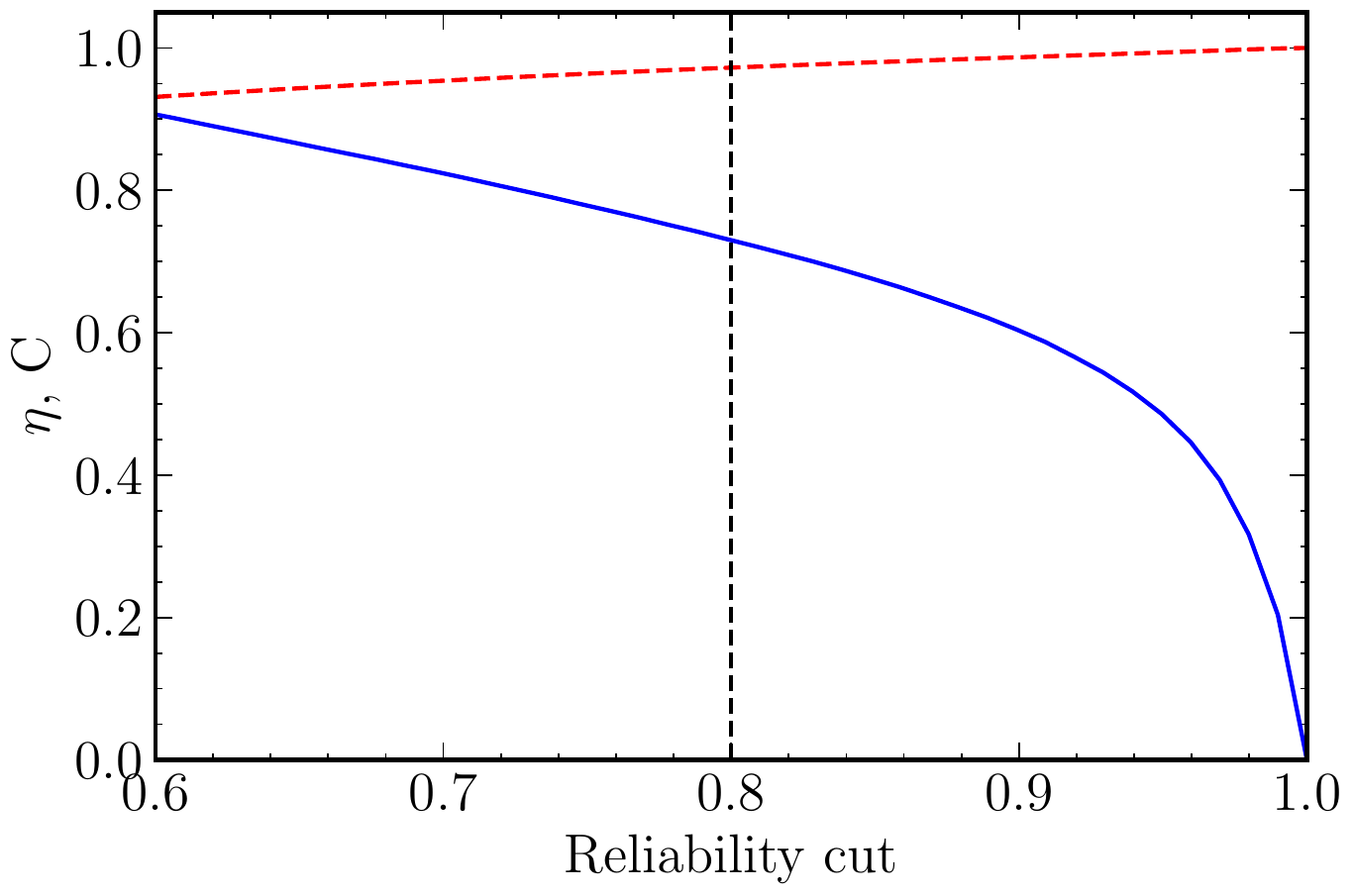}
	\caption{The completeness $\eta$ (solid blue line) and the cleanness C (dashed red line) of the counterpart sample as a function of reliability cut. The vertical dashed line shows the reliability cut used to define the reliable sample in this analysis.} 
	\label{fig:completeness_cleanness}
\end{figure}

The cleanness of the reliable sample is defined as

\begin{equation}
\label{eq:cleanness}
C = 1 - \frac{N_{\textrm{False}}}{N_{\textrm{SPIRE}}},
\end{equation}

\noindent which is illustrated alongside the completeness in Fig \ref{fig:completeness_cleanness}. The completeness and cleanness are plotted as functions of the minimum reliability that defines the sample. The figure shows that a reliability cut of $R \geq 0.8$ represents a good balance between high completeness and cleanness; a higher value of reliability would lead to a large drop in the completeness of the sample at little gain in cleanness and although a lower value would yield a much more complete sample, it would potentially lead to problems in ensuring a one-to-one matching of sources and counterparts. 

The LR method assumes that there is just one genuine counterpart associated with each source and forces sources with multiple candidates to have a sum of reliabilities that does not exceed 1. This means that for our reliability cut of R $\geq$ 0.8, any additional genuine counterparts, such as the second galaxy in a merging system, will reduce the value for both genuine counterparts to below our reliability threshold. We thus have a bias against multiple genuine counterparts such as merging galaxies or galaxies that form part of the same cluster. The presence of additional candidates in the search radius means that fewer IDs will be seen as reliable, leading to an increase in the incompleteness of the sample. In Table \ref{tab:multiplicity} we show the number of SPIRE sources and the fraction of those that are matched to a reliable counterpart as a function of the number of candidates found within 15 arcsec. The falling number in the fraction of reliable IDs is a sign of the incompleteness of the sample due to the increased number of potential candidates. We compare the effects of multiplicity with \citealt{Smith_2011}, \citealt{Fleuren_2012}, \citealt{Bourne_2016} and  \citealt{Furlanetto_2018} who all use search radii of 10 arcsec. The rate at which the reliable percentage decreases is much slower for our sample, which shows that a larger radius is in general more likely to detect a reliable counterpart. However, this also has the effect of increasing the likelihood that we match to an erroneous counterpart. The presence of many more candidates within the search radius reduces the reliability of the true ID and increases our false identification fraction. This may be part of the cause for the marginally higher false ID rate quoted above. We also note the low value for the percentage of reliable counterparts when there is only one possible candidate. A possible explanation for this also comes from the large choice of search radius. Using a radius of 15 arcsec means that occassionally the chosen ID will be a counterpart that is likely to be unassociated with the source and has a large angular separation from the SPIRE location, but is chosen regardless as it is the sole possible counterpart. To demonstrate this we have also shown the average separation between the chosen ID and the SPIRE location. The much higher value for when there is one counterpart suggests we should expect a lower reliable percentage. 

We repeated our Likelihood Ratio analysis using a 10 arcsec search radius to see if a reduction in the average number of candidates per source would lead to an increase in the total number of reliably matched sources and a decrease in the false ID rate. We found that at a search radius of 10 arcsec the LR method yielded only 421 more reliably matched \textit{Herschel} sources and the false ID rate decreased negligibly. Given the small difference between the two analyses, we preferentially chose to use a 15 arcsec search radius to avoid missing any unusual VIKING objects related to the \textit{Herschel} source. As shown in \citealp{Bakx_2020}, there are still genuine associations found on the VIKING images beyond 10 arcsec as a result of gravitational lensing. This may in part be the reason why our estimate of $Q_0$ continues to fall slowly between 10 and 15 arcsec. We wish to keep such VIKING objects in our catalogue despite them being highly unlikely to have significant reliability values.

\begin{table}
	\centering
	\caption{The number of SPIRE sources, those of which are denoted reliable and the average separation [arcsec] between the source and chosen ID as a function of the number of candidate IDs.}
	\label{tab:multiplicity}
	\begin{tabular}{lcccr}
		\hline
		\hline
		$N_{\textrm{match}}$ & $N_{\textrm{SPIRE}}$ & $N_{\textrm{Reliable}}$ &
		Per cent & Av. Separation \\
		\hline
		0 & 2 739 & 0 & 0 & 0 \\
		1 & 11 692 & 5 477 & 47 & 6.3 \\
		2 & 24 268 & 14 568 & 60 & 4.7 \\
		3 & 32 948 & 20 396 & 62 & 4.1 \\
		4 & 33 526 & 20 383 & 61 & 3.8 \\
		5 & 27 745 & 16 359 & 59 & 3.6 \\
		6 & 20 236 & 11 563 & 57 & 3.4 \\
		7 & 12 999 & 7 155 & 55 & 3.4 \\
		8 & 8 079 & 4 355 & 54 & 3.3 \\
		9 & 4 983 & 2 640 & 53 & 3.4 \\
		10 & 3 017 & 1 643 & 54 & 3.3 \\
		\hline
		\hline
	\end{tabular}
\end{table}

The number of missed genuine multiple counterparts can be estimated by assuming that all candidates are associated to a source if their combined reliability is greater than the 0.8 cut without any individual counterpart meeting the threshold (\citealt{Fleuren_2012}). However, due to the large search radius of 15 arcsec, the average number of candidates per SPIRE source is large meaning that even moderately valued counterparts will combine to exceed the reliability cut. Alternatively, we may consider using the LR values rather than the reliability. For sources where there is just one possible ID, equation (\ref{eq:reliability_multi}) tells us that an LR of 0.66 (assuming $Q_0 = 0.835$) corresponds to a reliability of 0.8. To mitigate the impact of multiple counterparts on the number of reliable IDs we may select those IDs that have an LR greater than 0.66 but fail to meet the reliability threshold. This set of IDs, though not included as part of the reliable sample, may be genuine multiple counterparts. We find 33 967 sources that satisfy these conditions, but note that these IDs will have a substantial fraction that are due to chance alignments. 

\begin{figure}
	\includegraphics[width=\columnwidth]{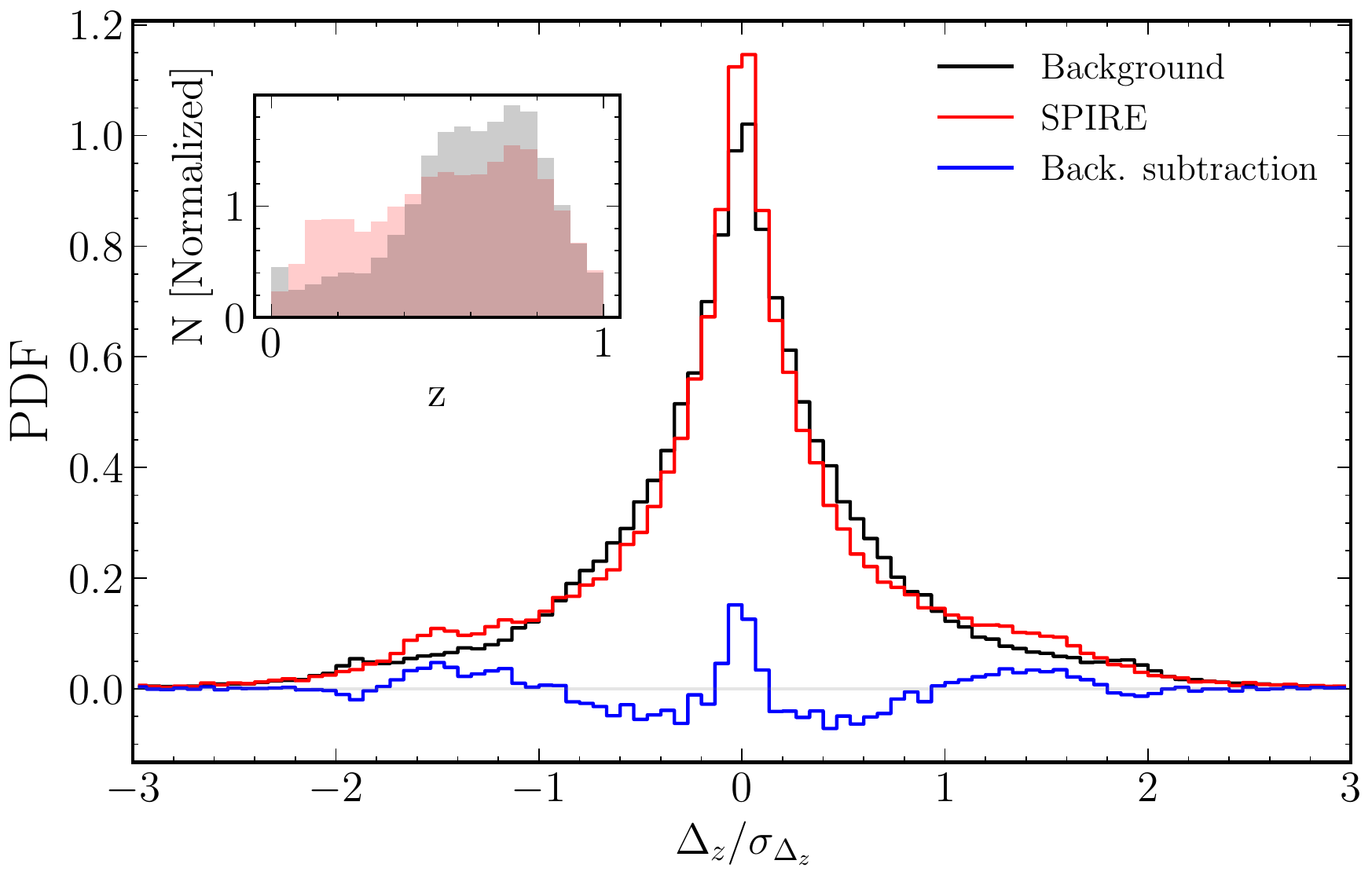}
	\caption{Histograms representing probability density functions for $\frac{\Delta_z}{\sigma_{\Delta_z}}$, the difference in photometric redshifts of counterparts divided by the error in $\Delta_z$ derived from the errors in redshifts. The red and black histograms represent the closest pair of counterparts found surrounding each \textit{Herschel} source and background position, respectively. The blue histogram is the result of subtracting the background from the \textit{Herschel} sources. The dashed blue line shows a Gaussian fit to this histogram, from which an estimate is made for the probability that a pair of VIKING counterparts is associated with the source. The inset figure shows the redshift distribution of VIKING counterparts found to within 8 arcsec of SPIRE (red histogram) and background (grey histogram) positions.} 
	\label{fig:deltaz}
\end{figure}

We can find a new estimate that removes chance alignments by considering the photometric redshifts of the VIKING objects (see Section \ref{sec:help_redshifts}). We first restrict our counterparts to those found within 8 arcsec, where we find the majority of L > 0.66 counterparts, reducing the chance of selecting multiple systems that are unrelated to the source. In Fig \ref{fig:deltaz} we show a probability density function (PDF) for the closest redshift pair of VIKING objects within 8 arcsec of all SPIRE and background positions. These distributions are given as a function of the difference in photometric redshifts, $\Delta_z$, divided by the error in $\Delta_z$. To estimate the number of genuine multiple counterparts that are a result of the \textit{Herschel} source we subtract the background distribution from our \textit{Herschel} candidates and fit a Gaussian profile to the excess found close to $\Delta_z = 0$ (blue histogram). The area of this peak represents an estimate for the probability that a close pair of VIKING galaxies observed within the search radius of a \textit{Herschel} position is indeed related to the source. Depending on bin width and the maximum allowed value of $\Delta_z/\sigma_{\Delta_z}$ we find this value to be in the range 2 -- 5\%. Multiplying this percentage by the number of sources that have a pair of candidates with $-0.1 \lesssim \Delta_z/\sigma_{\Delta_z} \lesssim 0.1$ yields an estimate for the number of \textit{Herschel} sources with genuine multiple counterparts of $\sim$ 400 -- 1 000. Beyond the central peak we see a well-defined shape to the excess function. This shape is the result of the differing redshift distributions of VIKING counterparts surrounding \textit{Herschel} positions and background positions to within 8 arcsec (see inset). The same differences are found for the distribution of errors on photometric redshifts.

This method of determining the number of possible interacting systems is independent of each counterpart’s reliability value and thus the closest redshift pair may not always coincide with the counterpart chosen by the Likelihood Ratio method. Of the 70 880 sources where there are $\geq$2 candidates with photometric redshifts within a search radius of 8 arcsec, approximately three quarters have counterparts that are matched by the LR method and are also part of the redshift pair. The redshift pairs of the remaining sources may be locations of unrelated interactions or multiples of the \textit{Herschel} source that are overlooked by the one-to-one matching of the LR method. Our list of 70 880 sources where we can estimate a value of $\Delta_z/\sigma_{\Delta_z}$ represents just over one third of all sources in the SGP, and so many more galaxy mergers or interactions may be in the catalogue. 

\subsection{Photometric Redshifts}
\label{sec:photometric_redshifts}

In this section we describe the method for obtaining redshift estimates for the sub-mm sources using their SPIRE flux densities and for the VIKING counterparts with aid of the Herschel Extragalactic Legacy Project (HELP; \citealp{Vaccari_2016}). 

\subsubsection{Herschel Redshifts}
\label{sec:herschel_redshifts}

We estimate a redshift for each \textit{Herschel} source by fitting a standard spectral energy distribution (SED) to the 250-, 350- and 500-$\mu$m flux densities and finding the redshift for which the following chi-squared sum reaches a minimum:

\begin{equation}
\label{eq:chi_square}
\chi^2 = \frac{(S_{250} - S_{250, m})^2}{E_{250}^2} + \frac{(S_{350} - S_{350, m})^2}{E_{350}^2} + \frac{(S_{500} - S_{500, m})^2}{E_{500}^2}.
\end{equation}

$S_{250}$, $S_{250,m}$ and $E_{250}$ represent the flux, model prediction of flux and the error at 250\,$\mu$m. We choose not to use the flux measurements from PACS at 100- and 160-$\mu$m as the \textit{Herschel} images are much less sensitive at these wavelengths compared to SPIRE. We use the SED derived by \citealt{Pearson_2013}, who use the flux densities from high redshift \textit{Herschel} sources that have spectroscopic redshifts to form a characteristic SED at high-z. The SED takes the form of a two temperature modified black-body with a cold dust temperature of 23.9\,K, a hot dust temperature of 46.9\,K and a ratio of cold dust mass to hot dust mass of 30.1. We vary the normalization and redshift of the model SED until the minimum chi-squared value is found and use this redshift as our estimate for the redshift of the source. The flux density measurements come from the H-ATLAS catalogues (\citealt{Valiante_2016}; \citealt{Maddox_2018}), but the flux errors listed in the catalogues do not include a calibration error. The calibration error consists of two parts: a 4\% error which is correlated between bands, caused by the uncertainty in the flux density measured for Neptune, and a 1.5\% error uncorrelated between bands (\citealt{Valiante_2016}). The correlated error measurement is not important to our analysis as the correlated error scales all flux densities by the same percentage and therefore does not contribute to the chi-squared sum.  The uncorrelated errors, however, do contribute to chi-squared, and thus the errors we used in the equation above come from adding in quadrature the errors in the catalogue with errors obtained by multiplying the flux density of each source by 0.015. For our investigation into gravitational lensing of sources it is also crucial that we have a reliable estimate on the error of our redshift values. To estimate the errors we find the redshifts above and below the best fit redshift at which the chi-squared increases by one. 

In such an analysis, the mean reduced chi-squared for the best fitting models for all the sources should have a value $\sim$ 1. A mean value >1 indicates that the model is not a good representation of the data and a mean value <1 implies that either the model has too many free parameters or that the errors on the data points are too large. Our initial analysis gave a mean chi-squared value of 0.55. In this case, the model is quite simple with only two free parameters: the amplitude of the SED and the redshift. The true explanation seems most likely to be that the errors are too large because there is an independent reason why this might be so. The errors in the H-ATLAS catalogues were produced by adding confusion noise and instrumental noise in quadrature, with confusion noise being slightly larger than the instrumental noise at all wavelengths \citep{Valiante_2016}. The confusion noise, however, will be strongly correlated between wavelengths and will therefore produce a smaller contribution to chi-squared than it would have if it were uncorrelated. On the assumption that this is the explanation of the low values of chi-squared, we reduced the errors by a factor of $\sqrt{0.55} = 0.74$ and repeated the SED fitting using these scaled errors.

These errors include the effect of the uncertainty in the flux densities but do not include the effect of the uncertainty in the model SED. The SED derived by \citealt{Pearson_2013} was the best fit to the flux densities of the sample of high-redshift \textit{Herschel} galaxies with spectroscopic redshifts, but \citealt{Bakx_2018} showed that high-redshift \textit{Herschel} galaxies have a range of SEDs. The sources in the sample of \citealt{Pearson_2013} are very bright and thus the errors in the redshifts due to the errors in the flux densities are small. Therefore, the redshift error estimated by \citealt{Pearson_2013}, $\Delta_z/(1+z) = 0.12$, is a good estimate of the fundamental redshift error that is caused by the diversity of real SEDs. For $\simeq$9\% of sources the method above produced a redshift error lower than this minimum value, and we have therefore replaced the redshift error from the chi-square fitting for these sources by this minimum redshift error.

\subsubsection{HELP Redshifts}
\label{sec:help_redshifts}

We obtain photometric redshift estimates for the near-infrared counterparts by matching our catalogue to the SGP data found in the Herschel Extragalactic Legacy Project (HELP). The method used in HELP to estimate redshifts come from the process outlined in \citealt{Duncan_2018a} and \citealt{Duncan_2018b}. The approach combines both SED template fitting and machine learning to generate multiple individual photometric estimates. Three different template sets are used to generate SED fitting based values: stellar-only templates from the default library of SEDs from the EAZY software (\citealt{Brammer_2008}); \citealt{Salvato_2009} XMM-COSMOS templates that include 30 SEDs covering a variety of galaxy spectral types and AGN and quasar templates; and \citealt{Brown_2014} Atlas of Galaxy SEDs, a set of 129 galaxy SED templates based on nearby galaxies that cover elliptical, spiral and luminous infrared galaxy spectral types. The filters used include $ugri$ from the OmegaCAM imager of the VLT Survey Telescope (VST), $griZY$ from the Dark Energy Camera (DECAM) and $ZYJHK_s$ from the VISTA InfraRed CAMera (VIRCAM). The machine learning photometric estimates are produced using the redshift code GPz (\citealt{Almosallam_2016a}; \citealt{Almosallam_2016b}), which uses Gaussian processes to map a given set of magnitudes and their uncertainties to spectroscopic redshifts. For the SGP, a spectroscopic sample of 48 995 galaxies plus additional extra redshifts from the GAMA fields formed the training set for the Gaussian process photometric estimates. This training set of spectroscopic redshifts come from the 2dF, 6dF, 2MRS and SRSS2 surveys. The sets of photo-z estimates are then combined using the Hierarchical Bayesian combination method which is described in \citealt{Dahlen_2013}.

We match with the resulting photometric sample for the SGP to find 542 302 matches to within 0.5 arcsec of the position of each VIKING galaxy, which results in 82 195 redshift estimates to our reliable (R $\geq$ 0.8) \textit{Herschel} sources list. This means that approximately 74 per cent of reliable sources have a redshift estimate for both the sub-mm source and the reliably matched candidate. A comparison of these redshifts is important in our investigation of gravitationally lensed sources so a sample of 74 per cent represents a large data set from which we can search for potentially lensed systems.

The Hierarchical Bayesian procedure used by HELP provides the full photo-z posterior for all sources in the HELP catalogue. For each redshift posterior, an 80\% highest probability density (HPD) credible interval (CI) is calculated by starting at the peak redshift probability and lowering a threshold until 80\% of the probability is included. The primary peak (and any potential secondary peak) is located by identifying the points where the posterior crosses this threshold. For each peak the median redshift within the boundaries of the 80\% HPD CI is used as estimates of the photometric redshift. We take as our estimate photo-z values to be the median redshift of the primary peak for each object matched to the HELP catalogue. We estimate errors on this value by taking the lower and upper boundaries of the HPD CI and transform them to one sigma errors assuming that the posterior is Gaussian in shape and symmetric about some mean redshift value such that an 80\% CI is equal to 1.282 $\sigma$.

\begin{figure}
	\includegraphics[width=\columnwidth]{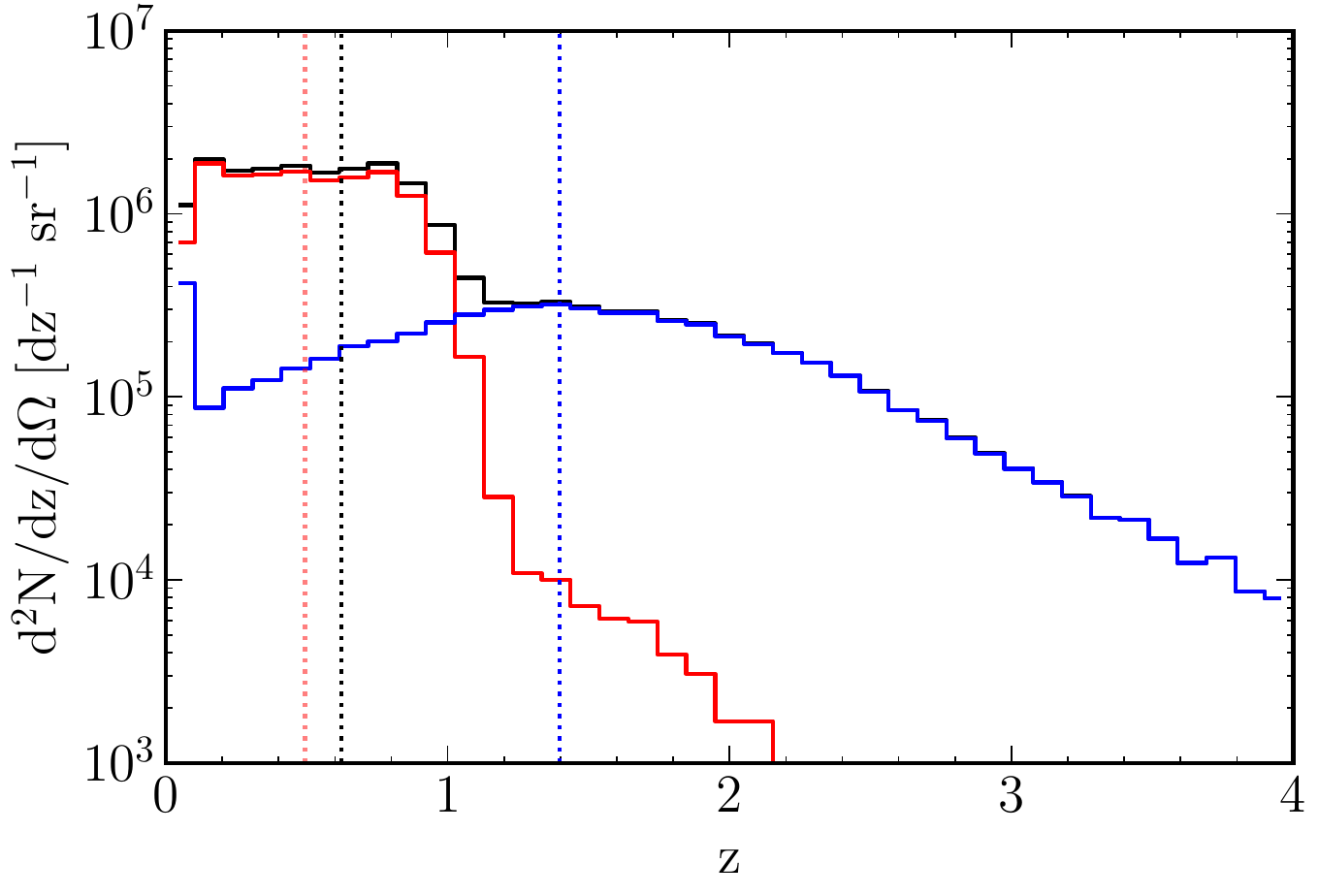}
	\caption{Photometric redshift distribution of the \textit{Herschel} sources. The red line represents the redshift distribution for counterparts that are matched to an object in the HELP catalogue. The blue line corresponds to sources that did not have a matched counterpart found in HELP, and thus uses the Herschel estimated redshift values. The black line represents the best estimate of photometric redshift for all sources. The dotted lines represent the median values for the three distributions: $z_{\textrm{best}} = 0.62$, $z_{\textrm{HELP}} = 0.49$ and $z_{\textrm{Herschel}} = 1.40$.} 
	\label{fig:z_distribution}
\end{figure}

The HELP photometric redshift estimates typically have smaller uncertainties than those estimated from the \textit{Herschel} flux densities, with the mean error on the \textit{Herschel} estimates being 0.51, compared to 0.21 for the counterparts matched to an object in HELP. For a full source catalogue of photometric redshifts we therefore preferentially choose the redshift estimate from HELP if one is available and when there is no match we rely on the value estimated by SPIRE flux densities. All three photo-z distributions can be seen in Fig \ref{fig:z_distribution}. Although it can be seen that the majority of our HELP matched counterparts have $z_{\textrm{phot}}$ < 1, which is to be expected given the limiting magnitude of VIKING, we note that this represents an improvement on the depth of galaxies observed with SDSS. Although our VIKING counterparts peak at $z \sim 0.5$, we find a substantial sample at larger redshifts and indeed at $z_{\textrm{phot}} \sim 1.5$ there are more than 200 HELP identifications. The \textit{Herschel} estimates have a much wider distribution with a median of $z_{\textrm{phot}} = 1.40$. Fig \ref{fig:z_distribution} shows an abundance of sources with high redshifts, including $\sim$ 100 that are estimated to be at greater than $z_{\textrm{phot}} \sim 5$. However, given that high redshift \textit{Herschel} sources are less likely to have counterparts on the VIKING images, the reliability decreases with increasing redshift from $\sim$ 0.9 for \textit{Herschel} redshift estimates between zero and one, down to $\sim$ 0.6 for redshift values greater than four. The choice of reliability cut will therefore greatly affect the depth of the sample.

\subsection{Source Counts}
\label{sec:source_counts}

\begin{figure*}
	\includegraphics[width=\textwidth]{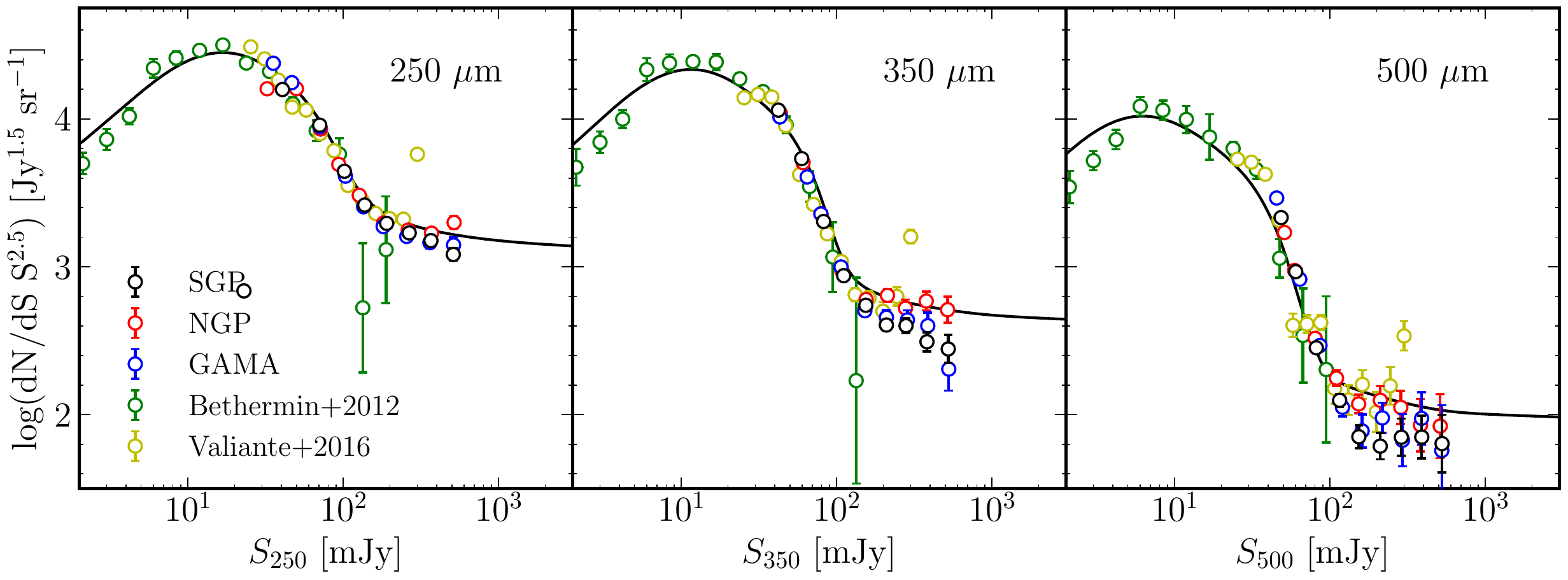}
	\caption{Euclidean normalized differential source counts for the three H-ATLAS fields at 250-, 350- and 500-$\mu$m compared to the galactic evolution model of \citealt{Cai_2013} (black line). The source counts for the GAMA fields (blue circles), the NGP (red circles) and the SGP (black circles) are shown alongside the counts estimated from \protect\citealt{Bethermin_2012} (green circles) and \protect\citealt{Valiante_2016} (yellow circles) for comparison. The errors on the data points are given by the square root of the raw counts in each flux density bin, apart from the errors for \protect\cite{Bethermin_2012} and \protect\citealt{Valiante_2016} where they are as quoted in their respective papers.} 
	\label{fig:model_comparison}
\end{figure*}

The Euclidean normalized number counts are derived from the data and shown in Fig \ref{fig:model_comparison} alongside calculations for the NGP and GAMA fields using the sources presented in DR1 and DR2 (\citealt{Valiante_2016} and \citealt{Maddox_2018}, respectively). We compare the counts to those estimated from the \textit{Herschel} Multi-tiered Extragalactic Survey (HerMES; \citealt{Oliver_2012}) of the COSMOS field by \citealt{Bethermin_2012} and the source counts for the GAMA fields that have been corrected for biasing effects using simulations in \citealt{Valiante_2016}. We also show the predictions from the galactic evolution model of \citealt{Cai_2013}\footnote{https://people.sissa.it/~zcai/} where the source counts reflect a maximum magnification due to gravitational lensing of $\mu$ = 12. The \citealt{Negrello_2007} model was the one that best fit the H-ATLAS source counts in our initial investigation of the source counts based on our results from the \textit{Herschel} SDP (\citealt{Clements_2010}). Here we choose to compare to the \citealt{Cai_2013} model which is a development of \citealt{Negrello_2007}. The "hybrid" model of \citealt{Cai_2013} combines a physical, forward model for protospheroidal galaxies and AGNs at z $\geq$ 1.5 with a phenomenological backward model for late-type galaxies and later AGN evolution. For the star-forming protospheroidal galaxies \citealt{Cai_2013} adopts the model of \citealt{Granato_2004}, which defines these high redshift sources as massive protospheroidal galaxies that are in the process of forming their stellar mass. The low redshift populations are split into "warm" starburst galaxies and "cold" normal late-type galaxies.

The counts in Fig \ref{fig:model_comparison} are characterized by a steep rise towards fainter fluxes starting at $\sim$ 100\,mJy. The steep rise is a signature of a strongly evolving luminous (> $10^{11} L_{\astrosun}$) population (\citealt{Clements_2010}) which, given its prominence at 350- and 500-$\mu$m compared to 250\,$\mu$m, indicates that this population is primarily at z > 1.5 (\citealt{Granato_2004}). The \citealt{Cai_2013} model provides a physical explanation for this rise. As the strongly negative K-correction at sub-mm wavelengths means that flux densities for a given luminosity are only weakly dependent on redshift and given that far-IR luminosity is approximately proportional to halo mass, the steep counts reflect the fall in the halo mass function at high masses. The change also implies a strong magnification bias as a result of gravitational lensing. The shape and location of this feature can therefore be used to put constraints on the evolution of this high redshift population.

The number counts for the GAMA, SGP and NGP fields are consistent with each other with the only noticable difference being between our recalculation of the GAMA source counts and the \citealt{Valiante_2016} counts at high flux densities. Fig \ref{fig:model_comparison} also shows an incompleteness to the source counts of \citealt{Bethermin_2012} at high flux densities. The \citealt{Cai_2013} model predicts a slightly higher normalization than those implied by our calculations for the three H-ATLAS regions. This small discrepancy for the asymptotic Euclidean counts at high flux suggests a difference in the expected number of sources in the region where local sub-mm galaxies are expected to reside. (For estimated values of the asymptotic Euclidean source count slopes at SPIRE wavelengths see \citealt{Serjeant_2005}). 

\section{Lensed Population}
\label{sec:lensed_population}

The magnification of sources due to the gravitational lensing effect of foreground objects makes searching for lensed systems an effective tool for studying the population of intrinsically faint and distant galaxies that lie below the \textit{Herschel} 500\,$\mu$m detection limit. These galaxies are prime targets for detailed study of the physical conditions one might find in distant dusty, star-forming galaxies and large samples of gravitational lenses provide huge potential as subjects for cosmological studies into measurements of cosmological parameters (e.g. \citealt{Kochanek_1992}; \citealt{Kochanek_1996}; \citealt{Grillo_2008}; \citealt{Oguri_2012}; \citealt{Eales_2015}) and the evolution of the equation of state of dark matter (e.g. \citealt{Zhang_2009}). 

It has been shown that wide area, blank-field submillimetre surveys provide an effective way of developing such large samples of gravitationally lensed, high-redshift galaxies using a selection of the brightest sources (e.g. \citealt{Blain_1996}; \citealt{Perrotta_2002}; \citealt{Negrello_2007}; \citealt{Paciga_2009}; \citealt{Bakx_2020}). Selecting lensed sources by means of high flux density cuts (typically > 100\,mJy at 500\,$\mu$m) has advantages not only in its ease of use but in the efficiency of the method and low contamination in the resulting sample. This simple selection criteria is expected to leave a set dominated by lensed sources as well as local late-type galaxies and blazars which can easily be removed by cross-matching with optical and radio surveys. Using an $S_{500}$ > 100\,mJy flux density cut \citealt{Negrello_2010} produced the first sample of 5 strongly lensed galaxies (SLGs) from the \textit{Herschel} SDP. Subsequently, \citealt{Wardlow_2013} identified 13 lensed galaxies over 95\,deg$^{2}$ of HerMES, \citealt{Nayyeri_2016} published a further 77 from the 372\,deg$^{2}$ of the HerMES Large Mode Survey (HeLMS) and \citealt{Negrello_2017} presented a sample of 80 candidate SLGs from 600\,deg$^{2}$ of the H-ATLAS.

\begin{figure}
	\includegraphics[width=\columnwidth]{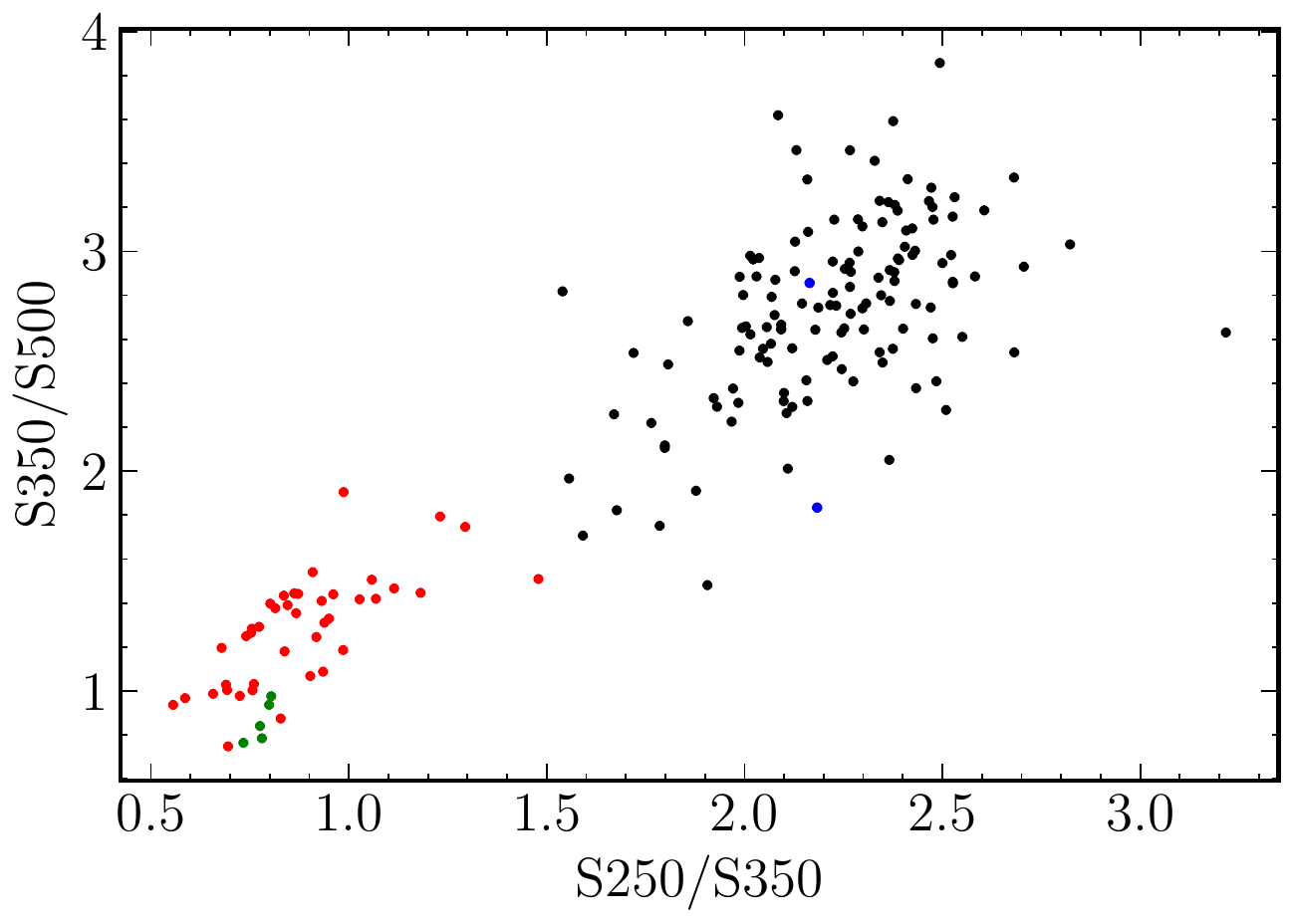}
	\caption{SPIRE colour-colour diagram for objects with $S_{500}$ > 100\,mJy. The colours represent the identification from NED: candidate lensed galaxies (red dots); local galaxies (black dots); stars (blue dots) and blazars (green dots).} 
	\label{fig:colour_100mjy}
\end{figure}

We present a sample of 41 candidate SLGs with flux density greater than 100\,mJy at 500\,$\mu$m from the SGP field, 30 of which form part of the \cite{Negrello_2017} sample and 11 are new candidates (see Table \ref{tab:SLG_candidates}). The 11 additional candidates fall outside the mask of the SGP used in the analysis of \citealt{Negrello_2017}, which was designed to remove the edges of the area observed by \textit{Herschel} that were in general affected by a higher noise. An initial selection of $S_{500}$ > 100\,mJy yields 179 sources with 175 of these matched to some VIKING counterpart. A fraction of these sources will be low-redshift spiral galaxies and flat spectrum radio galaxies that otherwise contaminate the set. We remove such objects by searching their location with the NASA/IPAC Extragalactic Database (NED). We identify 131 clear local galaxies with typically "bluer" sub-mm colours than candidate lensed galaxies. We also remove two variable stars HATLASJ012658.0-323234 (R Sculptoris) and HATLASJ225519.6-293644 (V Piscis Austrini) as well as five blazars (four of which were previously identified by \citealt{Negrello_2017}) that are listed in Table \ref{tab:blazars}. The remaining 41 sources are retained as our candidate lenses. Fig \ref{fig:colour_100mjy} shows a $S_{250}$/$S_{350}$ against $S_{350}$/$S_{500}$ colour-colour plot. As with the \citealt{Negrello_2017} sample, the lensed candidate galaxies occupy a "redder" proportion of the colour-colour space than those of local galaxies. The bimodality of the sample reflects the two populations being at very different redshifts, indicating that a relatively clean sample of lensed candidates (barring possible blazars) can be obtained from just submillimetre flux densities alone. This idea looks to be supported by the full \citealt{Negrello_2017} sample wherein only one candidate among the NGP and GAMA fields has been confirmed as not being a strongly lensed galaxy thus far. 

\begin{table*}
	\centering
	\caption{Blazars identified with $S_{500}$ > 100\,mJy.}
	\label{tab:blazars}
	\begin{tabular}{lcccr}
		\hline
		\hline
		H-ATLAS IAU name & NED Identification & $S_{250}$ (mJy) & $S_{350}$ (mJy) & $S_{500}$ (mJy)\\ 
		\hline
		HATLASJ014310.0-320056 & PKS 0140-322 & 96.0$\pm$7.5 & 119.5$\pm$8.4 & 122.4$\pm$9.0 \\
		HATLASJ014503.4-273333 & [HB89] 0142-278 & 131.4$\pm$7.8 & 179.2$\pm$8.8 & 234.4$\pm$9.0 \\
		HATLASJ222321.6-313701 & PKS 2220-318 & 86.0$\pm$9.5 & 110.9$\pm$10.5 & 131.9$\pm$11.7 \\
		HATLASJ224838.6-323551 & [HB89] 2245-328 & 119.2$\pm$7.7 & 152.8$\pm$8.3 & 194.7$\pm$8.6 \\
		HATLASJ235347.4-303746 & PKS 2351-309 & 77.1$\pm$7.4 & 96.6$\pm$8.4 & 103.1$\pm$8.9 \\
		\hline
		\hline
	\end{tabular}
\end{table*}

While we have presented evidence that these galaxies are at high redshift, we cannot say definitively that they are lensed. However, in the next section, we use a comparison of the estimated redshifts for the \textit{Herschel} sources and the redshifts of the near-IR counterparts to estimate the probability that these are indeed lensed. We also extend our analysis to much fainter sources to see whether it is practical to produce very large samples of lensed galaxies.

\subsection{In Search of Large Samples of Lensed Sources}
\label{sec:shalos_methodology}

One way to estimate the probability that a source is lensed, for the sources with a probable near-IR counterpart, is to compare the redshifts of the counterparts and the \textit{Herschel} source. If the redshift of the former is significantly lower, the source is likely to be lensed. In this section, we use an abbreviated version of the Statistical Herschel-ATLAS Lensed Objects Selection (SHALOS; \citealt{GonzalezNuevo_2019}) method to estimate the probability that the SGP sources are lensed.

The SHALOS analysis is a method for estimating the probability that a source is gravitationally lensed based on measures of similarity between the probability distributions of properties of the \textit{Herschel} source and the optical/near-IR counterpart, assuming one to be the source and the other being the foreground galaxy. To compare the similarity of two  probability distributions the method uses the Bhattacharyya distance (\citealt{Bhattacharyya_1943}), closely related to the Bhattacharyya coefficient, which estimates the overlap of the two distributions. For two continuous probability distribution functions $x$ and $y$ that have means $\mu_x$ and $\mu_y$ and standard deviations $\sigma_x$ and $\sigma_y$, respectively, the Bhattacharyya distance can be written as

\begin{equation}
\label{eq:bhattacharyya_distance}
D(x,y) = -\ln[BC(x,y)] = -\ln\Bigg[\int\dd{z}\sqrt{x(z)y(z)}\Bigg],
\end{equation}

\noindent where $BC(x, y)$ represents the Bhattacharyya coefficient. When the distributions $x$ and $y$ are assumed to be Gaussian, the distance measure can be calculated as:

\begin{equation}
\label{eq:bhattacharyya_distance_normal}
D(x, y) = \frac{1}{4}\ln\Bigg[\frac{1}{4}\Bigg(\frac{\sigma_x^2}{\sigma_y^2} + \frac{\sigma_y^2}{\sigma_x^2} + 2\Bigg)\Bigg] + \frac{1}{4}\Bigg[\frac{(\mu_x - \mu_y)^2}{\sigma_x^2 + \sigma_y^2}\Bigg]. 
\end{equation}

For a probability that is based on the redshift distributions of the source and counterpart, $p_z$, the system is most likely to represent a lensed system (with the near-infrared counterpart acting as the lens to the \textit{Herschel} source) when the redshift distributions are vastly different from each other. For this reason we use $1 - BC_z$ to assign probability values. In \citealt{GonzalezNuevo_2019} two further properties, the angular separation of the two objects on the sky and the optical to sub-mm flux density ratios, are also used to define additional probability values. These observables can be combined multiplicatively without losing generality or their meaning as probabilities to obtain a total probability of strong lensing for each candidate. In the case for \citealt{GonzalezNuevo_2019} this would be $p_{\textrm{total}} = BC_{\textrm{ang}} \ast (1 - BC_z) \ast (1 - BC_{\textrm{ratio}})$. \citealt{GonzalezNuevo_2019} also incorporate a fourth observable that does not rely on the Bhattacharyya coefficient based on the observed luminosity of the source. We implement just the redshift observable and use the probability derived from the comparison of the two redshift distributions as our lensing probability (the reader should therefore note that our sample and that of SHALOS are not comparable). We did not use the full SHALOS method for the following reason. A probability measure based on the angular separation would suggest that the closer the two objects are to the same line of sight, the more likely we would be observing a lensed system. This assumes that the lensing object must not have a large angular size, which limits us to galaxy-galaxy strong lensing events. As explained in \citealt{Bakx_2020}, the distribution of angular offsets between sources and optical/near-IR counterparts suggests that lensing by galaxy groups or clusters may contribute more significantly than galaxy-galaxy lensing for fainter flux densities. This has been illustrated by studies of the cross-correlation signal of foreground (GAMA and SDSS) and background (\textit{Herschel}) samples in \citealt{GonzalezNuevo_2014} and \citealt{GonzalezNuevo_2017}.

As the source needs to be at a higher redshift than the lens, we add to the lensing probability, $p_z$, any area of the source's redshift probability distribution function that lies below $\mu_{\textrm{VIKING}} - 3\sigma_{\textrm{VIKING}}$. Given that the VIKING photometric redshift errors are reltaively large (typically 50\%), this additional term is negligible for most sources. Only for sources where the VIKING redshift is estimated to be significantly higher than the Herschel source does this additional term have much bearing. As a result, no selection of sources with lensing probabilities above $\sim$ 0.6 will be affected by such contaminants. 

\subsection{Optimal Lensing Probability Threshold}
\label{sec:optimal_lensing_probability}

\begin{figure}
	\includegraphics[width=\columnwidth]{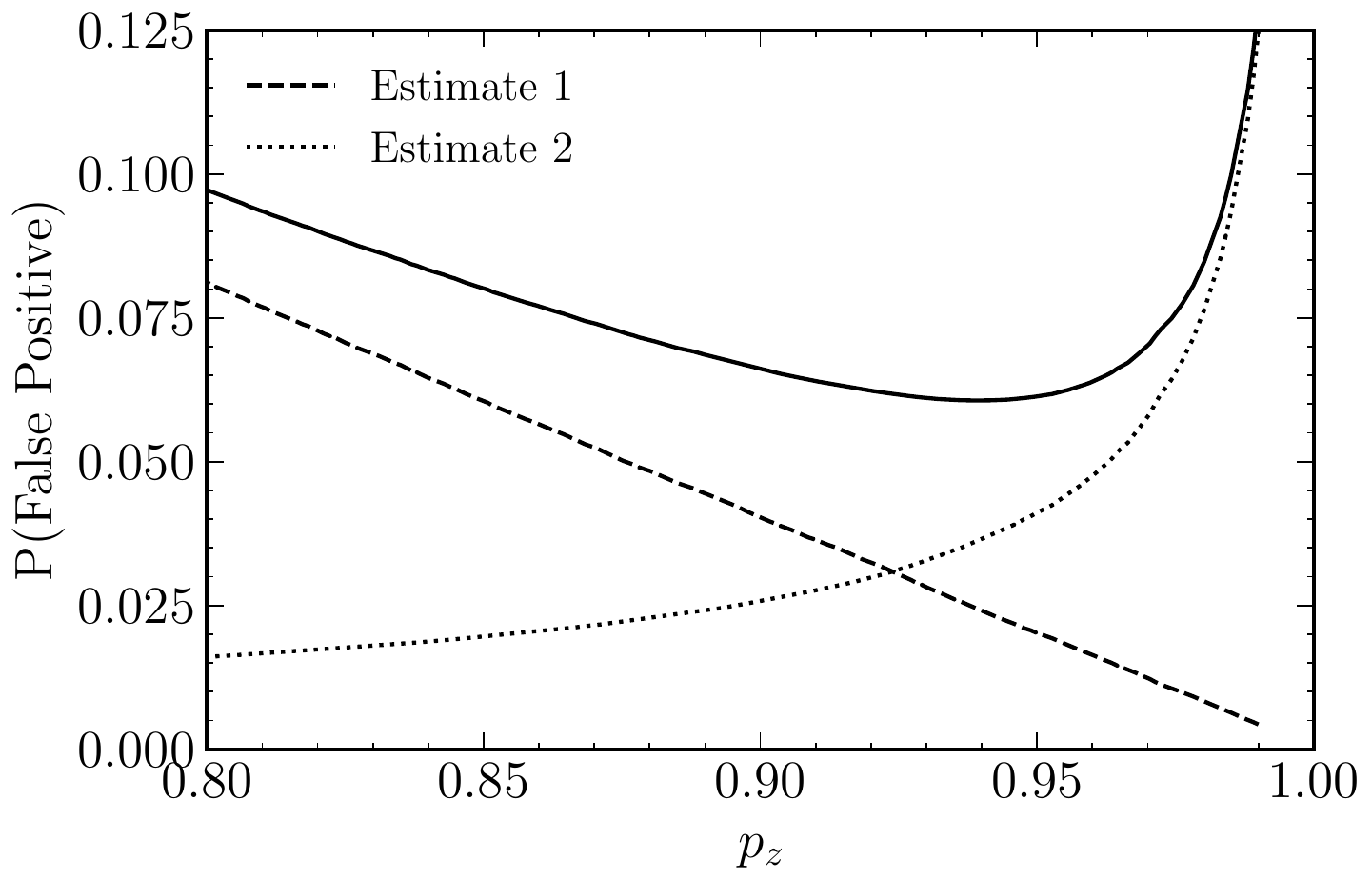}
	\caption{The probability of any given \textit{Herschel} source being unlensed within our lensed source catalogue, as a function of the lensing probability. This is estimated in two ways as detailed in the text: i) the sum $\Sigma_i^{N_{\textrm{lenses}}} (1 - p_{z,i})$ (dashed black line), ii) the fraction of sources in our reliable (R $\geq$ 0.8) catalogue that are deemed spurious and have a source redshift > 2.5, making it likely to be counted as a lensed source (dotted black line). The two estimates are combined for a total false positive rate (solid black line). The optimal lensing threshold is $p_{\textrm{optimal}} = 0.94$, where the false positive rate is estimated to be $\sim$ 6\%.} 
	\label{fig:false_positive_rate}
\end{figure}

Having estimated a value of $p_z$ for each \textit{Herschel} source based on the assumption that the VIKING counterpart is acting as a foreground deflector, we must decide on a threshold value, $p_{\textrm{critical}}$, for which a lensing probability greater than $p_{\textrm{critical}}$ is sufficiently high to be deemed part of our candidate lensed system catalogue. Our rationale for the choice of $p_{\textrm{critical}}$ is that it must construct a candidate sample that contains the fewest contaminating unlensed sources. While proving that any given \textit{Herschel} source within our catalogue is in fact not lensed is unattainable, we can estimate the number of contaminants we expect as a function of $p_z$. The value of $p_{\textrm{critical}}$ chosen is thus the value of $p_z$ that minimizes this false positive rate. We estimate this value in two ways. Firstly, the probability of any source, $i$, being unlensed is estimated as $1 - p_{z,i}$, so the sum

\begin{equation}
\label{eq:nunlensed_1}
N_{\textrm{unlensed, 1}} = \sum_i^{N_{\textrm{lenses}}} (1 - p_{z,i})
\end{equation}

\noindent for all sources above some value $p_z$ yields an estimate for the number of spuriously selected unlensed sources within a catalogue with minimum lensing probability $p_z$. Secondly, as mentioned in Section \ref{sec:reliable_id_sample}, our Likelihood Ratio analysis leads to a false identification rate of the reliable sample of 4.8\%. This means that for 4.8\% of sources with R $\geq$ 0.8, we expect there to be no association between the \textit{Herschel} source and the near-IR counterpart. We therefore make a second estimate that determines what fraction of all spuriously matched counterparts with R $\geq$ 0.8 would have been selected in our sample at a minimum threshold $p_z$. We calculate this value by taking the number of counterparts in our reliable sample, multiplying this by our false identification rate and multiplying again by the fraction of \textit{Herschel} sources at a high enough redshift for the near-IR and \textit{Herschel}/SPIRE redshifts to be signifcantly different (i.e., they are more likely to be classified as lensed than unlensed). We use a value of $z_{\textrm{Herschel}}$ > 2.5 as the lower limit for this final term, assuming a VIKING counterpart situated at $z_{\textrm{HELP}} \sim$ 0.5. The second estimate for the number of spurious unlensed sources can be written as:

\begin{equation}
\label{eq:nunlensed_2}
N_{\textrm{unlensed, 2}} = N_{\textrm{Reliable}} \times f_{\textrm{False}} \times {f_{\textrm{z}_{\textrm{Herschel}} > 2.5}}.
\end{equation}

Note that in using our false identifcation rate for reliably matched \textit{Herschel} sources, we make it a requirement that the \textit{Herschel} source has a VIKING counterpart with R $\geq$ 0.8. The two estimates are combined and divided by the total size of the unlensed catalogue to give a probability of any given source in our lensed catalogue being unlensed. This false positive rate is illustrated in Fig \ref{fig:false_positive_rate} as a function of $p_z$.

The figure shows that our summation of $1 - p_{z,i}$ method decreases steadily with increasing $p_z$. This trend is unsurprising; however, our second estimate shows a steady increase followed by a rapid rise in the number of false positives beyond $p_z \sim$ 0.95. This estimate reflects the number of \textit{Herschel} sources that will always appear unwantedly in our lensed catalogue. Therefore, as $p_z$ increases and the size of the lensed catalogue decreases, the fraction of false positives will rise. This, therefore, means that the second estimate will follow the general shape of the size of the catalogue with respect to $p_z$. Combining the two, we find a total rate that has a minimum at the optimal lensing probability threshold of $p_{\textrm{optimal}} = 0.94$. At this value the false positive rate is estimated to be $\sim 6\%$. Considering the criteria: i) a \textit{Herschel} source must have a VIKING counterpart with reliability $\geq$ 0.8, ii) the 500\,$\mu$m flux density must be < 100\,mJy (to avoid contamination with our manual search above this threshold), and iii) the lensing probability $p_z$ must be > 0.94, we estimate that the SGP contained 5 923 lensed sources.

In the catalogue we release with this paper we give values of $p_z$ for all sources with $p_z$ > 0.94. We note that the false positive rate for these lensed sources is about 6\%.

\subsection{Lensed Number Counts and Redshift Distribution}
\label{sec:lensed_number_counts}

\begin{figure}
	\includegraphics[width=\columnwidth]{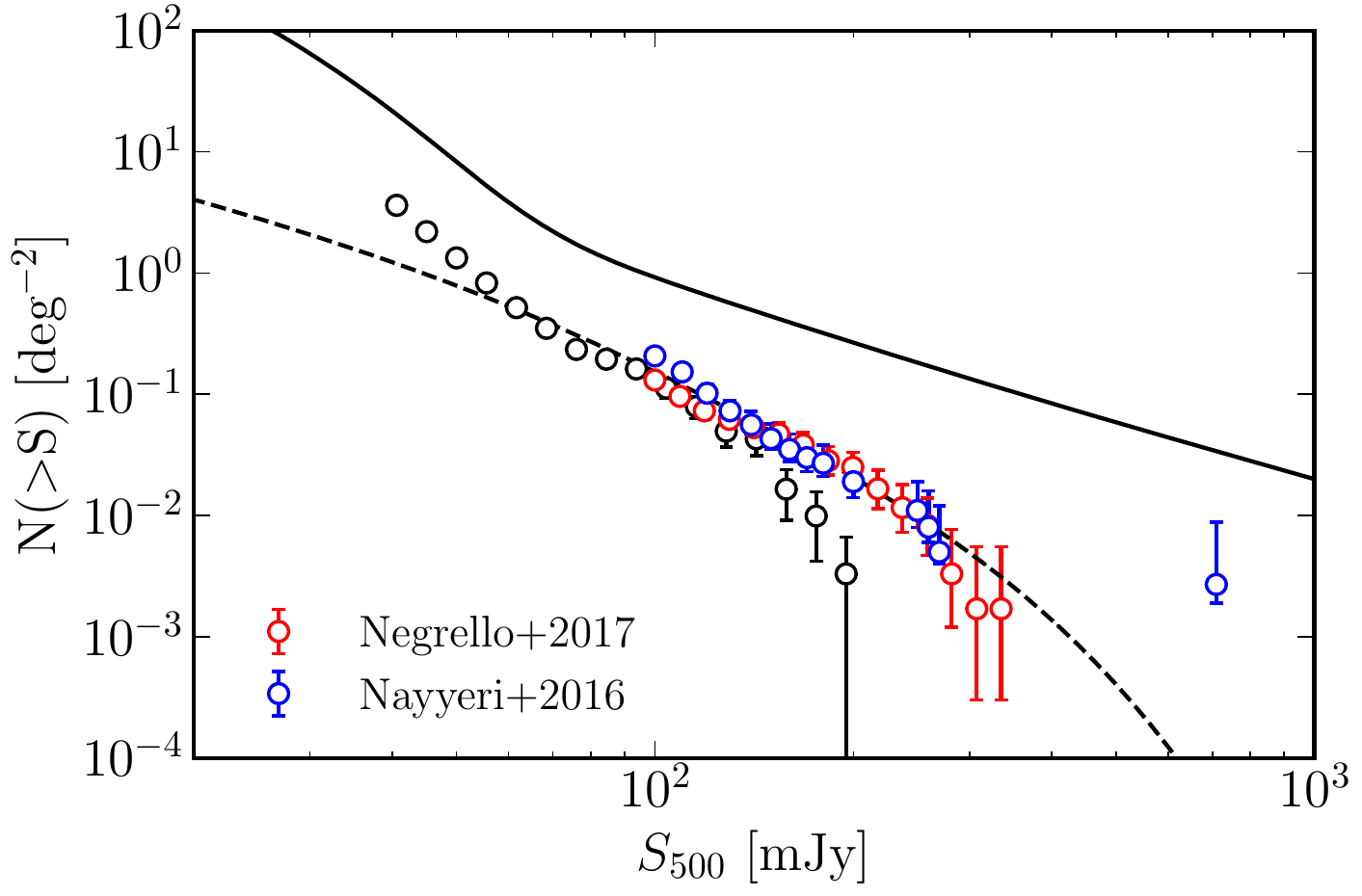}
	\caption{Cumulative number counts for lensed objects at 500\,$\mu$m compared to the predictions of the model of \protect\cite{Cai_2013} for lensed (dashed black line) and total source counts (solid black line). The counts and corresponding errors (square root of raw number counts in each flux density bin) for the predicted number of lensed sources are shown as black circles for a lensing probability of $p_{\textrm{z}}$ = $p_{\textrm{optimal}}$ = 0.94. The counts of candidate lensed sources from the \protect\citealt{Negrello_2007} and \protect\citealt{Nayyeri_2016} samples are shown as red and blue circles, respectively.} 
	\label{fig:lensed_number_counts}
\end{figure}

Fig \ref{fig:lensed_number_counts} shows the number counts for all the reliably matched SGP sources with $p_z > 0.94$ and $S_{500}$ < 100\,mJy, combined with the sources of Table \ref{tab:SLG_candidates}. We see that at high flux densities our sample falls below the prediction for the number of lensed systems made by \citealt{Cai_2013} (dashed black line) using a maximum magnification of $\mu$ = 12. \citealt{Negrello_2017} have already shown, however, that the number of lensed galaxies with $S_{500} \gtrsim$ 150\,mJy in the SGP is considerably lower than the other H-ATLAS fields and so the under prediction of our source counts is as expected. Above 100\,mJy, the surface density of strongly lensed galaxies has been found to be in the range $\sim$ 0.1 - 0.2\,deg$^{-2}$ (e.g. \citealt{Vieira_2010}; \citealt{Wardlow_2013}; \citealt{Nayyeri_2016}). Table \ref{tab:SLG_candidates} lists the number of high-z sources brighter than 100\,mJy which we have suggested might be lensed systems. 24 of these have near-IR counterparts with R $\geq$ 0.8. We can estimate values of $p_z$ for 19 of these, and in most cases these are > 90\%, additional evidence that these are lensed systems. The bright sources in Table \ref{tab:SLG_candidates} without near-IR counterparts might be interesting examples of sources in which the lens is at a very high redshift, might be being lensed by a cluster, or might not be lensed at all. If all the sources in Table \ref{tab:SLG_candidates} are being lensed, the surface density of lensed sources is 0.14\,deg$^{-2}$.

\begin{figure}
	\includegraphics[width=\columnwidth]{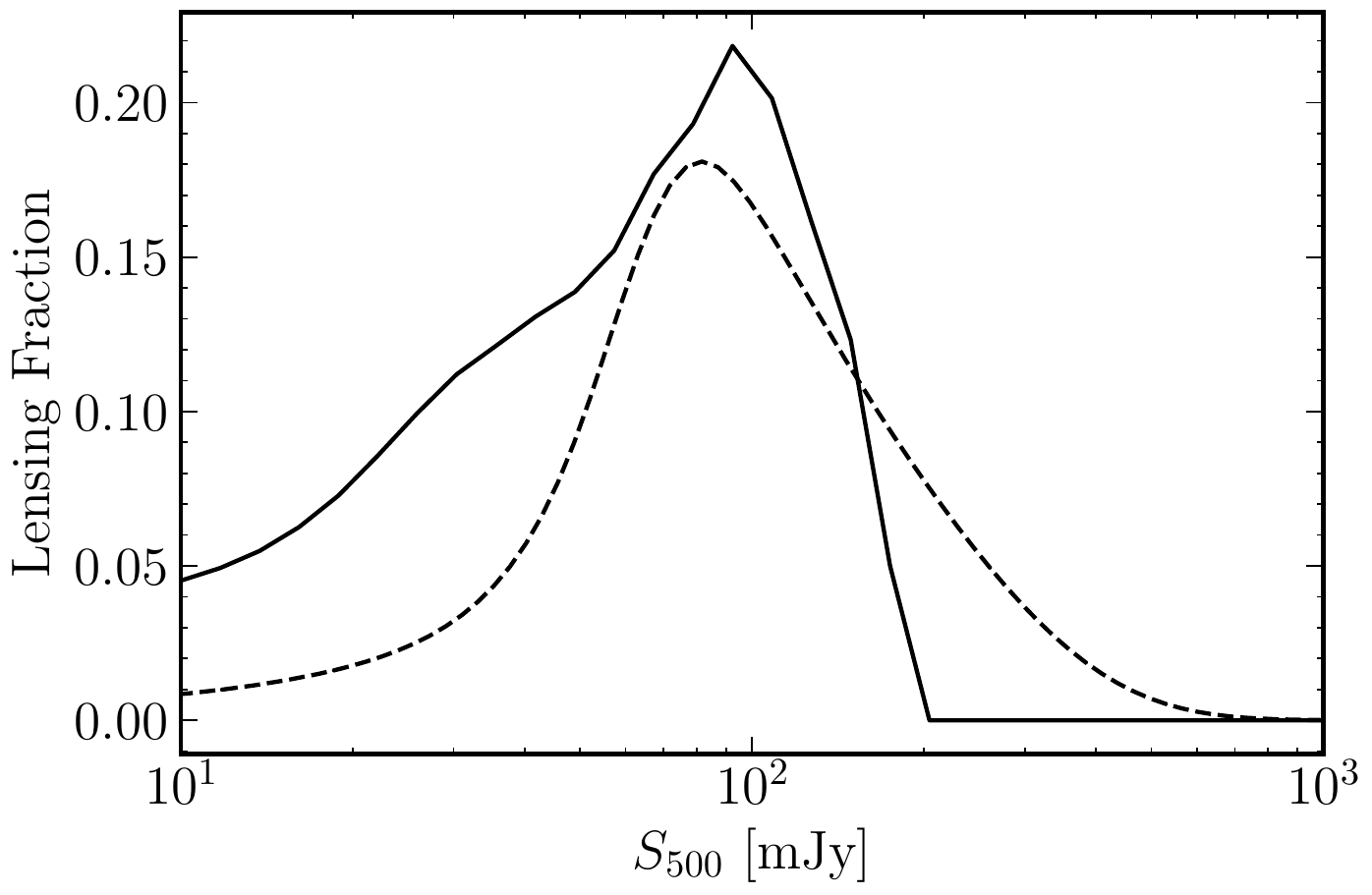}
	\caption{The fraction of lensed \textit{Herschel} sources with 500\,$\mu$m flux density greater than $S_{500}$. The solid black line represents a lensing probability of greater than 0.94 for sources with $S_{500}$ < 100\,mJy, combined with the lensing fraction estimated from Table \ref{tab:SLG_candidates} above this flux. The dashed black line represents the prediction from the \citealt{Cai_2013} model.}
	\label{fig:lensing_fraction}
\end{figure}

The fraction of all \textit{Herschel} sources that are part of our candidate lensed catalogue (at $p_{\textrm{optimal}} = 0.94$) is compared to the prediction of the \citealt{Cai_2013} model in Fig \ref{fig:lensing_fraction}. The model predicts that the fraction of sources that are gravitationally lensed rises rapidly from $\sim$ 50\,mJy and reaches a maximum by $\sim$ 80 -- 90\,mJy. Beyond this point the fraction falls as the brightest objects are dominated by local galaxies. Without such contaminants the lensing fraction would tend to one. At lower flux densities we find that the model predicts the lensing fraction to be very low below $\sim$ 40 -- 50\,mJy. Using a probability cut of 0.94 we predict a significantly higher lensing fraction than the model in this flux range. A possible explanation for this discrepancy may be the result of us comparing our simple method for detecting lenses with a model that is limited to strong lensing ($\mu$ > 2) events. We refer back to the weak lensing induced cross-correlation signal studies of \citealt{GonzalezNuevo_2014} and \citealt{GonzalezNuevo_2017}. These studies have shown that the cross-correlation signal between a foreground sample and background (\textit{Herschel}) sample has a sudden steepening at angular separations of $\sim$ 30 arcsec, indicating the typical angular scale where strong lensing becomes the dominant source of lensing. Our Likelihood Ratio search radius of 15 arcsec ensures that we are within this regime, and our criteria for reliably matched VIKING counterparts limits this further to typical angular separations no more than $\sim$ 8 arcsec (see Section \ref{sec:reliable_id_sample}). The inability to reproduce the observed cross-correlation at these small angular scales using a single halo model is interpreted by \citealt{GonzalezNuevo_2017} as being the result of additional strong lensing from the most massive foreground galaxies.

While simulations in \citealt{GonzalezNuevo_2014} fail to reproduce the signal solely by galaxy-galaxy weak lensing, it can be reproduced on sub-arcminute scales if the SDSS/GAMA foreground sample of galaxies act as signposts for galaxy groups or clusters. Our excess observed in the number counts at low flux densities may therefore be attributed to a collection of strongly and weakly lensed systems. We expect the counterparts closest to their respective \textit{Herschel} source to be producing the highest amplification factors, while the others may represent small magnifications of the \textit{Herschel} sources due to weak lensing, where the VIKING galaxies mark the locations of galaxy groups or clusters.

\begin{figure}
	\includegraphics[width=\columnwidth]{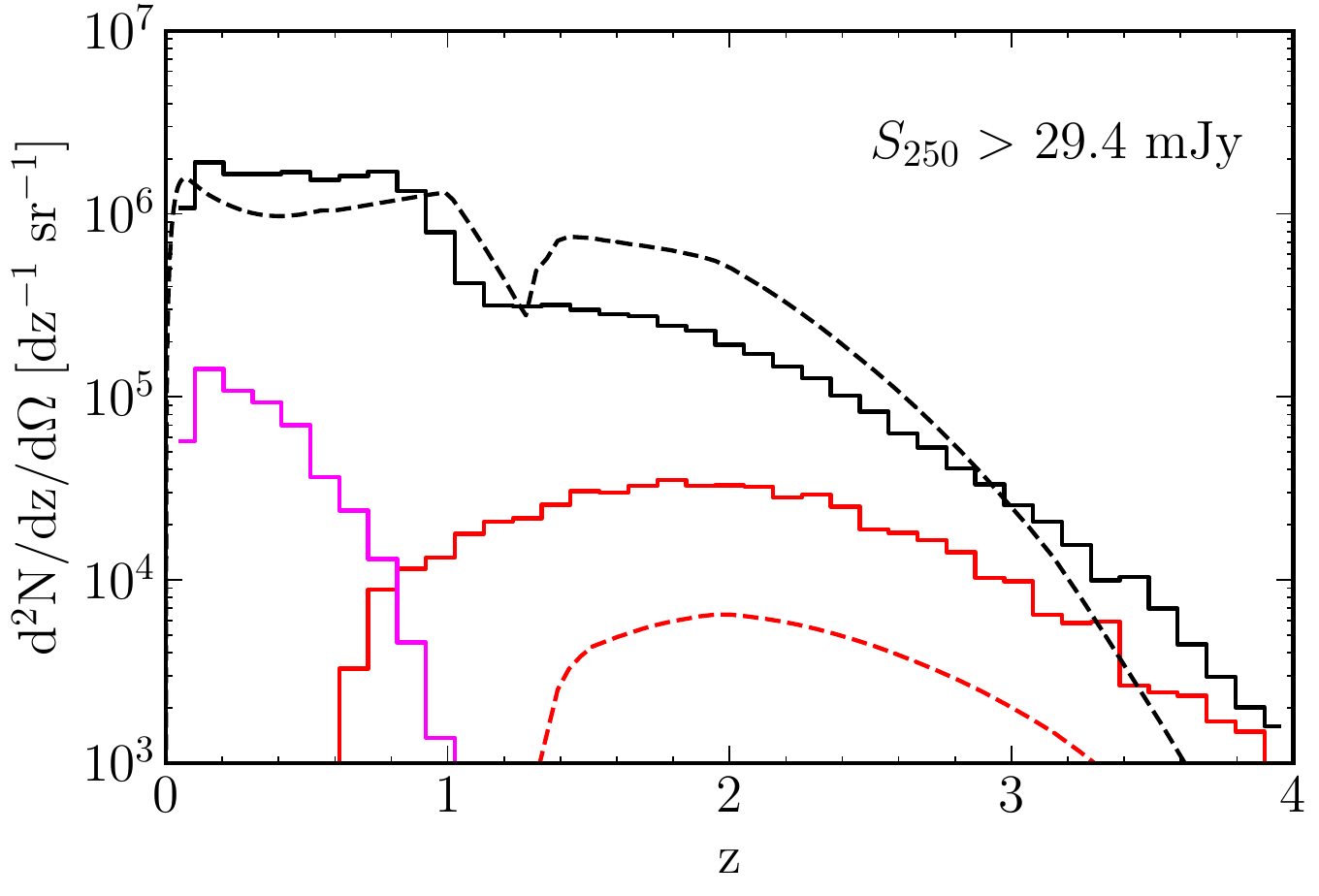}
	\caption{Photometric redshift distribution of \textit{Herschel} sources with 250\,$\mu$m flux density greater than 29.4\,mJy. The solid black line represents the best estimate of photometric redshift for all sources. The red solid line shows the distribution of \textit{Herschel} estimated redshifts for lensed candidates with a lensing probability greater than 0.94. The magenta solid line represents the redshift distribution for the same sources but for the HELP redshift estimates that are likely to represent the lenses. The median values for the three distributions are: $z_{\textrm{best}} = 0.62$, $z_{\textrm{Lensed (Source)}} = 2.03$ and $z_{\textrm{Lensed (Lens)}} = 0.29$. The black and red dashed lines illustrate the prediction from \citealt{Cai_2013} for the redshift distribution of all sources and lensed sources, respectively.} 
	\label{fig:lens_z_distribution}
\end{figure}

The redshift distribution for our sample of candidate lensed systems is shown in Fig \ref{fig:lens_z_distribution}. We have used a lensing probability cut of 0.94 and compared the distributions to the predictions of the \citealt{Cai_2013} model. We see that the model agrees well with the redshift distribution for all sources (black lines) but is an underestimate for the distribution of lensed sources (red lines; the magenta line represents the distribution of possible deflectors). As above, a possible explanation for this difference may be due to the effects of weak lensing that are not considered in the \citealt{Cai_2013} model.

\subsection{Comparison with \textit{Herschel}-ATLAS Lensed Objects Selection}
\label{sec:lensed_redshift}

It is possible to select a sample of SLGs from the SGP region using only information gained from the sub-mm sources, as described in the Herschel-ATLAS Lensed Objects Selection (HALOS; \citealt{GonzalezNuevo_2012}). This method selects candidates below the 100\,mJy flux limit, reaching a surface density of $\sim$ 1.5 -- 2\,deg$^{-2}$, with an expected efficiency of $\sim$ 50\%. Using the HALOS criteria: $S_{350} \geq 85$\,mJy, $S_{250} \geq 35$\,mJy, $S_{350}/S_{250} > 0.6$ and $S_{500}/S_{350} > 0.4$, we select 654 candidates, 392 of which are candidates for which we have calculated lensing probabilities. These sources lie predominantly at high lensing probabilities; 162 having a value of 0.94 or greater. \citealt{GonzalezNuevo_2012} continues by showing that the selection efficiency can be further improved by considering those sources that dominate the bright end of the high redshift far-IR/sub-mm luminosity function. \citealt{GonzalezNuevo_2012} estimate that objects with 100\,$\mu$m luminosity in the top 2\% of sources may generate a sample of SLG candidates with an efficiency of $\sim$ 70\% (based on a comparison with the HALOS method that uses information from VIKING counterparts). An investigation into the luminosity function calculated from our candidate sample would provide a good test for the expected efficiency of our method.

\section{Conclusions}
\label{sec:conclusions}

In this paper we described the process of identifying reliable near-infrared counterparts to sub-mm sources in the \textit{Herschel} DR2 across the SGP. The data described in this paper, added to the analyses of previous H-ATLAS fields, effectively doubles the number of reliably matched \textit{Herschel} sources now available.

We employ the likelihood ratio method to determine the reliability of all possible VIKING counterparts within 15 arcsec of 250\,$\mu$m selected sources. We found reliable (R $\geq$ 0.8) counterparts for 111 065 sub-mm sources, which corresponds to a return of 57.4 per cent of all sources. We estimate a false identification rate of 4.8 per cent, marginally higher than the result found by previous H-ATLAS analyses (e.g. \citealt{Fleuren_2012}; \citealt{Bourne_2016}; \citealt{Furlanetto_2018}), which can be attributed to the larger search radius used and the choice to keep sources with detections $\geq$ 4$\sigma$ at 350- and 500-$\mu$m. Using a set of background locations, we estimate the fraction of \textit{Herschel} sources that have a counterpart in the VIKING survey to be $Q_0 = 0.835\pm0.009$ ($0.823\pm0.009$ for extragalactic objects), which is very close to the value found by \citealt{Furlanetto_2018} for the NGP, but higher than the value given in \citealt{Fleuren_2012} which also uses VIKING near-infrared data. By looking at sources where there were several potential near-IR IDs and then comparing the estimated redshifts of these counterparts, we estimate that $\sim$ 400 -- 1 000 of the sources in the SGP have multiple genuine identifications, probably caused by galaxy mergers or interactions.

We determine the number counts of sub-mm sources in the 250-, 350- and 500-$\mu$m SPIRE bands and compare these observations to the galactic evolution model of \citealt{Cai_2013}. The observations show a steep rise in the counts at flux densities $\sim$ 100\,mJy which becomes most prominent at 350- and 500-$\mu$m, suggesting the presence of a strongly evolving luminous population at high redshift. We find the \citealt{Cai_2013} model accurately predicts the steepness of the counts at different wavelengths and thus gives an indication that the physical approach to the evolution of high redshift, star forming proto-spheroidal galaxies in this model is reasonably well founded.

We search for a large sample of lensed sources in the SGP by estimating for each source a probability that it is gravitationally lensed. This probability is defined by comparing the redshifts of the near-IR counterpart and the \textit{Herschel} source, as is described in the SHALOS method of \citealt{GonzalezNuevo_2019}. Assuming a probability cut of 0.94, we estimate that 5 923 \textit{Herschel} sources in the SGP are lensed with a lensing false positive rate of about 6\%. Above 100\,mJy at 500\,$\mu$m the incompleteness in matching to the Herschel Extragalactic Legacy Project for photometric redshift estimates of VIKING counterparts leads to missing lensing probabilities for the brightest, and thus most probable, lensed high redshift sources. We therefore include in Table \ref{tab:SLG_candidates} 41 high-z sources brighter than 100\,mJy that we suggest may be gravitationally lensed. This list has already been screened for local galaxies and flat spectrum radio galaxies that may also be above the 100\,mJy threshold. If all the candidates in Table \ref{tab:SLG_candidates} are lensed, then we predict the surface density of lensed sources to be 0.14\,deg$^{-2}$.

This data release provides an important update to the existing \textit{Herschel}-ATLAS catalogue, further enhancing the number of matched sources in the largest \textit{Herschel} survey. The full H-ATLAS now includes over 400 000 galaxies (with almost half of these being reliably matched to optical or near-infrared counterparts) spanning out to redshifts of 6. This makes it the most comprehensive unbiased sample of galaxies available for the study of obscured star formation and the evolution of the dusty interstellar medium to date. 

\section*{Acknowledgements}

We would like to thank the many astronomers who have contributed over the last decade to the \textit{Herschel}-ATLAS project. SAE and MWLS acknowledge the funding from the UK Science and Technology Facilities Council consolidated grant ST/K000926/1. LD, SJM and MWLS acknowledge the funding from the European Research Council (ERC) in the form of Consolidator Grant COSMICDUST (ERC-2014-CoG-647939).

\section*{Data Availability}

The data underlying this article are available at the H-ATLAS webpage; \url{www.h-atlas.org}.

\bibliographystyle{mnras}
\bibliography{bibliography}

\appendix

\section{Candidate lensed galaxies}

We present the 41 candidate lensed galaxies with $S_{500}$ > 100 mJy. 

\begin{table*}
	\centering
	\caption{Candidate lensed galaxies with $S_{500}$ > 100\,mJy. Each source has the following information: H-ATLAS IAU name, 250-, 350- and 500-$\mu$m flux densities, photometric redshift estimates from HELP, photometric redshift estimates from \textit{Herschel} flux densities and lensing probability.}
	\begin{threeparttable}
	\label{tab:SLG_candidates}
	\begin{tabular}{lcccccr}
		\hline
		\hline
		H-ATLAS IAU name & $S_{250}$ (mJy) & $S_{350}$ (mJy) & $S_{500}$ (mJy) & $z_{\textrm{HELP}}$ & $z_{\textrm{Herschel}}$ & $p_{\textrm{z}}$ \\
		\hline
		HATLASJ000007.5-334100 & 130.3$\pm$7.6 & 160.0$\pm$8.4 & 116.3$\pm$8.4 & 0.44$\pm$0.19 & 2.50$\pm$0.42 & 0.994 \\
		HATLASJ000124.9-354212\tnote{\textdagger} & 63.3$\pm$7.9 & 91.1$\pm$8.4 & 121.7$\pm$8.9 & - & 4.56$\pm$0.67 & - \\
		HATLASJ000722.2-352015 & 237.4$\pm$7.5 & 192.9$\pm$8.2 & 107.5$\pm$8.6 & 0.34$\pm$0.13 & 1.46$\pm$0.30 & 0.960 \\
		HATLASJ000912.7-300807 & 352.8$\pm$7.6 & 272.6$\pm$8.6 & 156.1$\pm$8.7 & 0.28$\pm$0.08 & 1.40$\pm$0.29 & 0.977 \\
		HATLASJ001010.5-360237\tnote{\textdagger} & 137.8$\pm$9.9 & 163.0$\pm$10.0 & 117.2$\pm$10.8 & - & 2.42$\pm$0.41 & - \\
		HATLASJ002624.8-341738 & 137.7$\pm$7.5 & 185.9$\pm$8.3 & 148.8$\pm$8.7 & 0.93$\pm$0.35 & 2.87$\pm$0.46 & 0.940 \\
		HATLASJ003207.7-303724 & 80.3$\pm$7.3 & 106.1$\pm$7.9 & 105.8$\pm$8.3 & 0.69$\pm$0.29 & 3.39$\pm$0.53 & 0.994 \\ 
		HATLASJ004736.0-272951 & 170.9$\pm$7.5 & 197.1$\pm$8.5 & 145.6$\pm$8.9 & 0.57$\pm$0.28 & 2.42$\pm$0.41 & 0.971 \\ 
		HATLASJ004853.3-303110 & 118.1$\pm$6.9 & 147.3$\pm$7.8 & 105.4$\pm$8.1 & 0.51$\pm$0.22 & 2.51$\pm$0.42 & 0.989 \\ 
		HATLASJ005132.0-302012 & 119.3$\pm$7.3 & 121.0$\pm$8.3 & 102.0$\pm$8.5 & 0.33$\pm$0.12 & 2.40$\pm$0.41 & 0.998 \\
		HATLASJ005132.8-301848 & 164.6$\pm$7.6 & 160.2$\pm$8.4 & 113.1$\pm$9.1 & 0.88$\pm$0.45 & 2.03$\pm$0.36 & 0.633 \\
		HATLASJ005724.2-273122 & 73.3$\pm$7.6 & 101.2$\pm$8.4 & 103.6$\pm$9.0 & 0.89$\pm$0.41 & 3.57$\pm$0.55 & 0.980 \\
		HATLASJ010250.9-311723 & 267.9$\pm$7.5 & 253.2$\pm$8.3 & 168.1$\pm$8.9 & 0.83$\pm$0.39 & 1.90$\pm$0.35 & 0.649 \\
		HATLASJ011424.0-333614 & 72.2$\pm$7.2 & 129.8$\pm$8.0 & 138.6$\pm$8.6 & 0.46$\pm$0.16 & 4.42$\pm$0.65 & 1.000$^{\ast}$ \\
		HATLASJ011947.1-272408 & 226.2$\pm$8.0 & 152.9$\pm$8.8 & 101.3$\pm$9.5 & 0.94$\pm$0.40 & 1.27$\pm$0.27 & 0.139 \\
		HATLASJ012046.5-282403 & 103.3$\pm$7.8 & 149.8$\pm$8.3 & 145.7$\pm$9.2 & 0.43$\pm$0.15 & 3.54$\pm$0.54 & 1.000$^{\ast}$ \\
		HATLASJ012407.4-281434 & 257.5$\pm$8.1 & 271.1$\pm$8.5 & 204.0$\pm$8.7 & 0.71$\pm$0.28 & 2.28$\pm$0.39 & 0.932 \\
		HATLASJ012416.0-310500 & 140.4$\pm$7.6 & 154.5$\pm$8.3 & 100.3$\pm$8.8 & 0.44$\pm$0.19 & 2.14$\pm$0.38 & 0.984 \\
		HATLASJ013004.1-305514 & 164.4$\pm$6.8 & 147.5$\pm$7.9 & 100.6$\pm$8.0 & 0.88$\pm$0.41 & 1.83$\pm$0.34 & 0.549 \\
		HATLASJ013240.0-330907 & 112.0$\pm$7.6 & 148.8$\pm$8.7 & 117.7$\pm$8.8 & - & 2.82$\pm$0.46 & - \\
		HATLASJ013840.5-281856 & 116.3$\pm$7.9 & 177.0$\pm$8.5 & 179.3$\pm$9.0 & 0.61$\pm$0.28 & 3.79$\pm$0.57 & 0.998 \\
		HATLASJ013951.9-321446 & 109.0$\pm$7.2 & 116.5$\pm$8.0 & 107.1$\pm$8.3 & 0.86$\pm$0.43 & 2.70$\pm$0.44 & 0.888 \\
		HATLASJ014849.3-331820 & 124.8$\pm$9.8 & 149.4$\pm$10.3 & 104.2$\pm$11.0 & 1.11$\pm$0.52 & 2.39$\pm$0.41 & 0.614 \\
		HATLASJ222536.3-295649 & 208.9$\pm$9.4 & 211.7$\pm$11.8 & 111.1$\pm$11.8 & - & 1.78$\pm$0.33 & - \\
		HATLASJ223753.8-305828 & 139.1$\pm$7.2 & 144.8$\pm$7.9 & 100.6$\pm$8.3 & - & 2.13$\pm$0.38 & - \\
		HATLASJ223955.2-290917 & 163.1$\pm$8.5 & 152.7$\pm$9.7 & 107.6$\pm$11.8 & - & 1.94$\pm$0.35 & - \\
		HATLASJ224207.2-324159 & 73.0$\pm$7.7 & 88.1$\pm$8.6 & 100.8$\pm$9.4 & - & 3.57$\pm$0.55 & - \\
		HATLASJ224805.4-335820 & 122.3$\pm$7.8 & 135.6$\pm$8.7 & 126.9$\pm$9.0 & - & 2.82$\pm$0.46 & - \\
		HATLASJ225250.7-313658 & 127.4$\pm$6.7 & 138.7$\pm$7.7 & 111.4$\pm$8.0 & - & 2.45$\pm$0.41 & - \\
		HATLASJ225844.8-295125 & 175.4$\pm$7.4 & 186.9$\pm$8.4 & 142.6$\pm$9.3 & 0.69$\pm$0.27 & 2.32$\pm$0.40 & 0.946 \\
		HATLASJ230546.3-331039 & 76.8$\pm$7.7 & 110.9$\pm$8.4 & 110.4$\pm$8.8 & 0.60$\pm$0.13 & 3.60$\pm$0.55 & 0.999 \\
		HATLASJ230815.6-343801 & 79.4$\pm$7.6 & 135.4$\pm$8.3 & 140.0$\pm$8.9 & 0.72$\pm$0.21 & 4.16$\pm$0.62 & 0.999 \\
		HATLASJ232419.8-323927 & 212.9$\pm$6.8 & 244.2$\pm$7.6 & 169.4$\pm$7.9 & 0.75$\pm$0.28 & 2.30$\pm$0.40 & 0.925 \\
		HATLASJ232531.4-302236 & 175.6$\pm$6.8 & 227.0$\pm$7.6 & 175.7$\pm$7.9 & - & 2.72$\pm$0.45 & - \\
		HATLASJ232623.0-342642 & 153.7$\pm$6.9 & 178.4$\pm$7.8 & 123.5$\pm$8.2 & - & 2.33$\pm$0.40 & - \\
		HATLASJ232900.6-321744 & 118.3$\pm$7.1 & 141.3$\pm$7.9 & 119.7$\pm$8.4 & 0.77$\pm$0.29 & 2.75$\pm$0.45 & 0.968 \\
		HATLASJ233720.9-293023 & 155.1$\pm$8.4 & 166.5$\pm$8.9 & 118.1$\pm$9.7 & - & 2.22$\pm$0.39 & - \\
		HATLASJ234357.7-351724 & 263.5$\pm$7.5 & 223.1$\pm$8.3 & 154.2$\pm$8.8 & - & 1.74$\pm$0.33 & - \\
		HATLASJ234418.1-303936 & 125.8$\pm$7.4 & 185.5$\pm$8.2 & 155.1$\pm$8.9 & 0.90$\pm$0.27 & 3.16$\pm$0.50 & 0.982 \\
		HATLASJ235623.1-354119 & 121.5$\pm$7.8 & 161.0$\pm$8.8 & 125.5$\pm$9.1 & - & 2.78$\pm$0.45 & - \\
		HATLASJ235827.7-323244 & 112.5$\pm$7.0 & 148.0$\pm$7.9 & 143.4$\pm$8.1 & - & 3.30$\pm$0.52 & - \\
		\hline
		\hline
	\end{tabular}
	\begin{tablenotes}
		\item[*] Lensing probability equal to one due to rounding.
		\item[\textdagger] No matched counterpart found.
	\end{tablenotes}
	\end{threeparttable}
\end{table*}

\bsp	%
\label{lastpage}
\end{document}